\documentclass[11pt]{article}
\usepackage[dvips]{epsfig}
\usepackage{delarray}
\usepackage{url}
\usepackage{anysize}
\usepackage{multirow}
\usepackage{amsmath}
\usepackage{graphicx}
\usepackage[OT1]{fontenc}
\usepackage{amsthm,amsmath}
\usepackage{color}
\usepackage[dvips]{epsfig}
\usepackage{delarray}
\usepackage{url}
\usepackage{colortbl}
\usepackage{anysize}
\usepackage{multirow}
\usepackage{amsmath}
\usepackage{amsthm}
\usepackage{amsmath, amssymb}
\numberwithin{equation}{section}
\newtheorem{theorem}{Theorem}

\newcommand{\etal}{{\em et al.}}

\newcommand{\bb}{\mbox{\bf b}}
\newcommand{\bff}{\mbox{\bf f}}
\newcommand{\bx}{\mbox{\bf x}}
\newcommand{\by}{\mbox{\bf y}}
\newcommand{\bA}{\mbox{\bf A}}
\newcommand{\ba}{\mbox{\bf a}}
\newcommand{\bB}{\mbox{\bf B}}

\newcommand{\bI}{\mbox{\bf I}}

\newcommand{\bR}{\mbox{\bf R}}
\newcommand{\bS}{\mbox{\bf S}}
\newcommand{\bX}{\mbox{\bf X}}
\newcommand{\bY}{\mbox{\bf Y}}
\newcommand{\bone}{\mbox{\bf 1}}
\newcommand{\bsone}{\mbox{\scriptsize \bf 1}}
\newcommand{\bzero}{\mbox{\bf 0}}
\newcommand{\bveps}{\mbox{\boldmath $\varepsilon$}}

\newcommand{\bxi}{\mbox{\boldmath $\xi$}}
\newcommand{\beps}{\mbox{\boldmath $\varepsilon$}}
\newcommand{\bmu}{\mbox{\boldmath $\mu$}}
\newcommand{\bgamma}{\mbox{\boldmath $\gamma$}}

\newcommand{\bSigma}{\mbox{\boldmath $\Sigma$}}

\newcommand{\cov}{\mathrm{cov}}
\newcommand{\Sig}{\mathbf{\Sigma}}
\newcommand{\veps}{\varepsilon}

\newcommand{\diag}{\mathrm{diag}}
\newcommand{\argmin}{\mathrm{argmin}}

\newcommand{\bw}{\mbox{\bf w}}
\newcommand{\var}{\mathrm{var}}

\newcommand{\calC}{{\cal C}}

\marginsize{1 in}{1 in}{1 in}{1 in}

\renewcommand{\baselinestretch}{1.3} 

\makeatletter

\def\singlespace{\def\baselinestretch{1}\@normalsize}

\newcounter{sect}  \newcounter{subsect}
 \newcounter{subsubsect}

\newcounter{prop}

\title{\bf Asset Allocation and Risk Assessment with Gross Exposure Constraints
for Vast Portfolios
\thanks{We thank conference participants of the
``15th Annual Conference on Derivative Securities and Risk
Management" at Columbia University, ``Financial Econometrics and
Vast Data Conference'' at Oxford University, ``The fourth
Cambridge-Princeton Conference in Finance'' at Cambridge University,
``The 2nd international conference on risk management'' in
Singapore, ``The 2008 International Symposium on Financial
Engineering and Risk Management'' in Shanghai, and seminar
participants of University of Chicago for helpful comments and
suggestions.   The financial support from NSF grants DMS-0532370 and
DMS-0704337 are greatly acknowledged. \textbf{Address Information:}
Jianqing Fan, Bendheim Center for Finance, Princeton University,
26 Prospect Avenue, Princeton, NJ 08540, USA. 
E-mail: \texttt{jqfan@princeton.edu}. Jingjin Zhang and Ke Yu,
Department of Operations Research and Financial Engineering,
Princeton University, Princeton, NJ 08540. E-mail:
\texttt{jingjinz@Princeton.edu}, \texttt{kyu@Princeton.edu}. } }
\author{Jianqing Fan, Jingjin Zhang and Ke Yu\\ Princeton University}

\date{\today}

\begin{document}

\maketitle

\begin{singlespace}
\begin{abstract}
Markowitz (1952, 1959) laid down the ground-breaking work on the
mean-variance analysis.  Under his framework, the theoretical
optimal allocation vector can be very different from the estimated
one for large portfolios due to the intrinsic difficulty of
estimating a vast covariance matrix and return vector. This can
result in adverse performance in portfolio selected based on
empirical data due to the accumulation of estimation errors. We
address this problem by introducing the gross-exposure constrained
mean-variance portfolio selection. We show that with gross-exposure
constraint the theoretical optimal portfolios have similar
performance to the empirically selected ones based on estimated
covariance matrices and there is no error accumulation effect from
estimation of vast covariance matrices. This gives theoretical
justification to the empirical results in Jagannathan and Ma (2003).
We also show that the no-short-sale portfolio is not diversified
enough and can be improved by allowing some short positions. As the
constraint on short sales relaxes, the number of selected assets
varies from a small number to the total number of stocks, when
tracking portfolios or selecting assets. This achieves the optimal
sparse portfolio selection, which has close performance to the
theoretical optimal one. Among 1000 stocks, for example, we are able
to identify all optimal subsets of portfolios of different sizes,
their associated allocation vectors, and their estimated risks. The
utility of our new approach is illustrated by simulation and
empirical studies on the 100 Fama-French industrial portfolios and
the 400 stocks randomly selected from Russell 3000.

\end{abstract}
\end{singlespace}

\newpage

\section{Introduction}

Portfolio selection and optimization has been a fundamental problem
in finance ever since Markowitz (1952, 1959) laid down the
ground-breaking work on the mean-variance analysis. Markowitz posed
the mean-variance analysis by solving a quadratic optimization
problem. This approach has had a profound impact on the financial
economics and is a milestone of modern finance. It leads to the
celebrated Capital Asset Pricing Model (CAPM), developed by Sharpe
(1964), Lintner (1965) and Black (1972). However, there are
documented facts that the Markowitz portfolio is very sensitive to
errors in the estimates of the inputs, namely the expected return
and the covariance matrix. One of the problems is the computational
difficulty associated with solving a large-scale quadratic
optimization problem with a dense covariance matrix (Konno and
Hiroaki, 1991). Green and Hollified (1992) argued that the presence
of a dominant factor would result in extreme negative weights in
mean-variance efficient portfolios even in the absence of the
estimation errors. Chopra and Ziemba (1993) showed that small
changes in the input parameters can result in large changes in the
optimal portfolio allocation.  Laloux \etal (1999) found that
Markowitz's portfolio optimization based on a sample covariance
matrix is not adequate because its lowest eigenvalues associated
with the smallest risk portfolio are dominated by estimation noise.
These problems get more pronounced when the portfolio size is large.
In fact, Jagannathan and Ma (2003) showed that optimal no-short-sale
portfolio outperforms the Markowitz portfolio, when the portfolio
size is large.

To appreciate the challenge of dimensionality, suppose that we have
2,000 stocks to be allocated or managed.  The covariance matrix
alone involves over 2,000,000 unknown parameters.  Yet, the sample
size $n$ is usually no more than 400 (about two-year daily data, or
eight-year weekly data, or thirty-year monthly data).  Now, each
element in the covariance matrix is estimated with the accuracy of
order $O(n^{-\frac{1}{2}})$ or 0.05.  Aggregating them over millions
of estimates in the covariance matrix can lead to devastating
effects, which can result in adverse performance in the selected
portfolio based on empirical data. As a result, the allocation
vector that we get based on the empirical data can be very different
from the allocation vector we want based on the theoretical inputs.
Hence, the mean-variance optimal portfolio does not perform well in
empirical applications, and it is very important to find a robust
portfolio that does not depend on the aggregation of estimation
errors.

Several techniques have been suggested to reduce the sensitivity of
the Markowitz-optimal portfolios to input uncertainty. Chopra and
Ziemba (1993) proposed a James-Stein estimator for the means and
Ledoit and Wolf (2003, 2004) proposed to shrink a covariance matrix
towards either the identity matrix or the covariance matrix implied
by the factor structure, while Klein and Bawa (1976) and Frost and
Savarino (1986) suggested Bayesian estimation of means and
covariance matrix. Fan \etal (2008) studied the covariance matrix
estimated based on the factor model and demonstrated that the
resulting allocation vector significantly outperforms the allocation
vector based on the sample covariance. Pesaran and Zaffaroni (2008)
investigated how the optimal allocation vector depends on the
covariance matrix with a factor structure when portfolio size is
large.  However, these techniques, while reducing the sensitivity of
input vectors in the mean-variance allocation, are not enough to
address the adverse effect due to the accumulation of estimation
errors, particularly when portfolio size is large. Some of
theoretical results on this aspect have been unveiled by Fan \etal
(2008).

Various efforts have been made to modify the Markowitz unconstrained
mean-variance optimization problem to make the resulting allocation
depend less sensitively on the input vectors such as the expected
returns and covariance matrices.  De Roon \etal (2001) considered
testing-variance spanning with the no-short-sale constraint.
Goldfarb and Iyengar (2003) studied some robust portfolio selection
problems that make allocation vectors less sensitive to the input
vectors. The seminal paper by Jagannathan and Ma (2003) imposed the
no-short-sale constraint on the Markowitz mean-variance optimization
problem and gave insightful explanation and demonstration of why the
constraints help even when they are wrong. They demonstrated that
their constrained efficient portfolio problem is equivalent to the
Markowitz problem with covariance estimated by the maximum
likelihood estimate with the same constraint.  However, as
demonstrated in this paper, the optimal no-short-sale portfolio is
not diversified enough.  The constraint on gross exposure needs
relaxing in order to enlarge the pools of admissible
portfolios.\footnote{Independently, DeMiguel \etal (2008), Bordie
\etal (2008) and this paper all extended the work by Jagannathan and
Ma (2003) by relaxing the gross-exposure constraint, with very
different focuses and studies. DeMiguel \etal (2008) focuses on the
effect of the constraint on the covariance regularization, a
technical extension of the result in Jagannathan and Ma (2003).
Bordie \etal (2008) and this paper emphasize on the sparsity of the
portfolio allocation and optimization algorithms. A prominent
contribution of this paper is to provide mathematical insights to
the utility approximations with the gross-exposure constraint.} We
will provide statistical insights to the question why the constraint
on gross exposure prevents the risks or utilities of selected
portfolios from accumulation of statistical estimation errors.  This
is a prominent contribution of this paper in addition to the
utilities of our formulation in portfolio selection, tracking, and
improvement.  Our result provides a thoeretical insight to the
phenomenon, observed by Jagannathan and Ma (2003), why the wrong
constraint helps on risk reduction for large portfolios.

We approach the utility optimization problem by introducing a
gross-exposure constraint on the allocation vector.  This makes not
only the Markowitz problem more practical, but also bridges the gap
between the no-short-sale utility optimization problem of
Jagannathan and Ma (2003) and the unconstrained utility optimization
problem of Markowitz (1952, 1959).  As the gross exposure parameter
relaxes from 1 to infinity, our utility optimization progressively
ranges from no short-sale constraint to no constraint on short
sales.  We will demonstrate that for a wide range of the constraint
parameters, the optimal portfolio does not sensitively depend on the
estimation errors of the input vectors. The theoretical (oracle)
optimal portfolio and empirical optimal portfolio have approximately
the same utility. In addition, the empirical and theoretical risks
are also approximately the same for any allocation vector satisfying
the gross-exposure constraint. The extent to which the
gross-exposure constraint impacts on utility approximations is
explicitly unveiled. These theoretical results are demonstrated by
several simulation and empirical studies. They lend further support
to the conclusions made by Jagannathan and Ma (2003) in their
empirical studies.

To better appreciate the above arguments, the actual risk of a
portfolio selected based on the empirical data can be decomposed
into two parts:  the actual risk (oracle risk) of the theoretically
optimal portfolio constructed from the true covariance matrix and
the approximation error, which is the difference between the two. As
the gross-exposure constraint relaxes, the oracle risk decreases.
When the theoretical portfolio reaches certain size, the marginal
gain by including more assets is vanishing.  On the other hand, the
risk approximation error grows quickly when the exposure parameter
is large for vast portfolios.  The cost can quickly exceed the
benefit of relaxing the gross-exposure constraint.  The risk
approximation error is maximized when there is no constraint on the
gross-exposure and this can easily exceed its benefit.   On the
other hand, the risk approximation error is minimized for the
no-short-sale portfolio, and this can exceed the cost due to the
constraint.

\begin{figure}[t]   
\begin{center}
\begin{tabular}{c c}
   \includegraphics[scale=0.5]{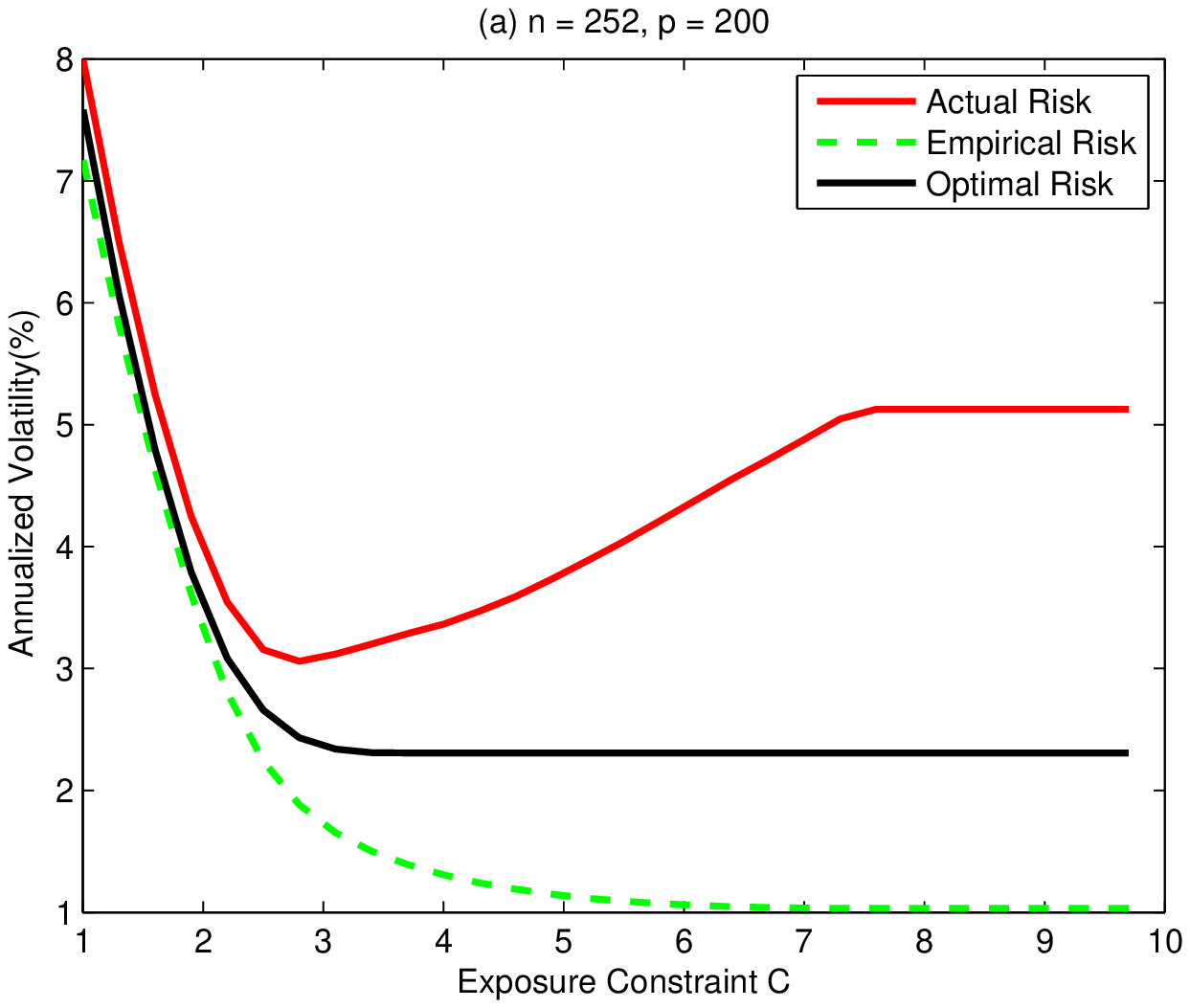} &
   \includegraphics[scale=0.5]{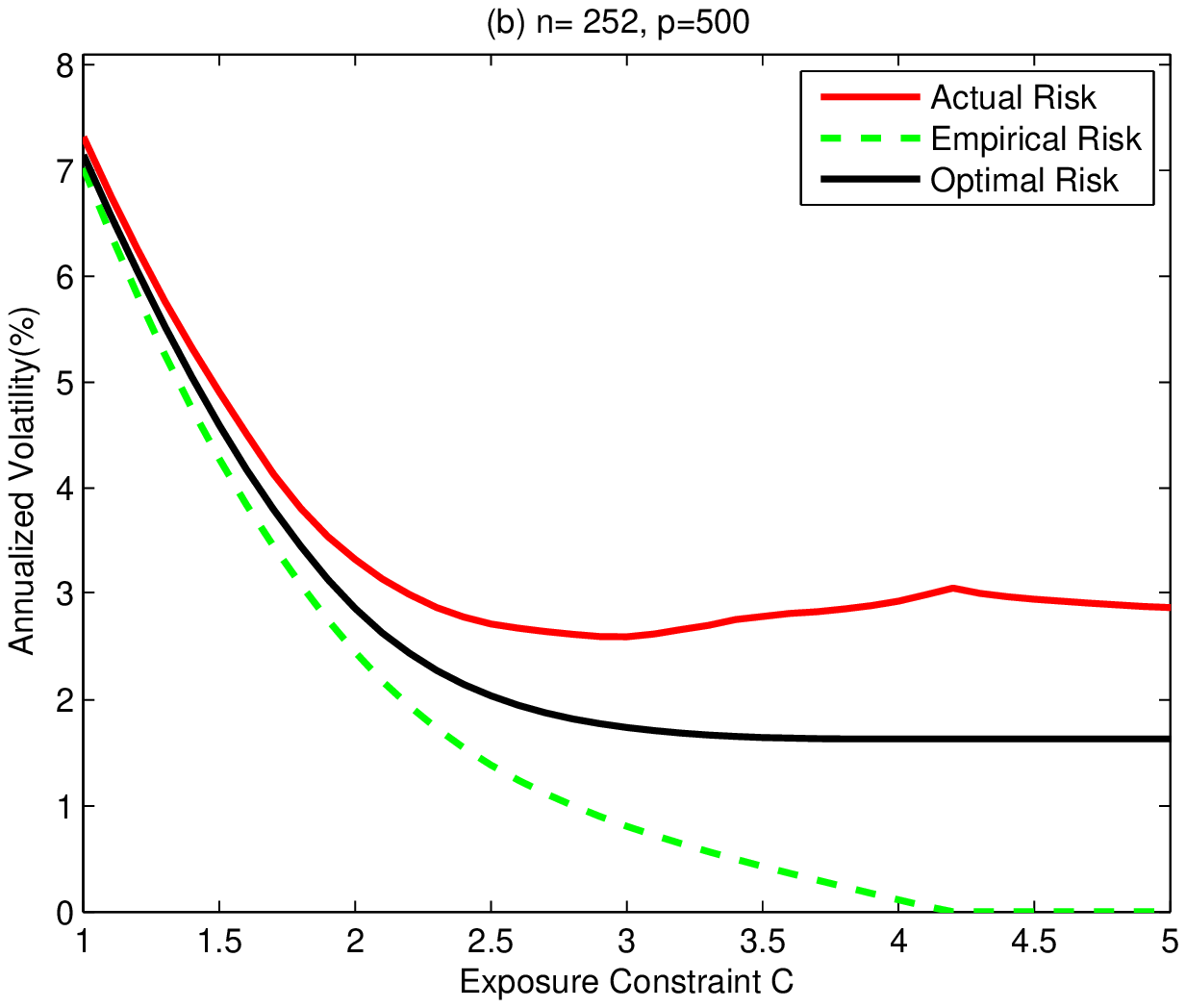}

\end{tabular}
\begin{singlespace}
   \caption{The risks of theoretically optimal portfolios, and the
   actual risks of the empirically optimal portfolios, and the empirical risks of
   the empirically optimal portfolios under gross-exposure
   constraints are plotted against the gross-exposure parameter $c$.  The data are
   based simulated 252 daily returns of 500 stocks from
   the Fama-French three-factor
   model.  As the gross-exposure parameter $c$
increases, the discrepancy between the optimal risks, actual risks,
empirical risks  get larger, which means the actual risk might be
quite far away from the risk we think it should be.  The total
number of stocks under consideration is (a) 200 and (b) 500.}
\end{singlespace}
\end{center}
\end{figure}

The above arguments can be better appreciated by using Figure 1, in
which 252 daily returns for 500 stocks were generated from the
Fama-French three-factor model, detailed in Section 4.  We use the
simulated data, instead of the empirical data, as we know the actual
risks in the simulated model. The risks of optimal portfolios stop
to decreases further when the gross exposure constant $c \geq 3$. On
the other hand, based on the sample covariance matrix, one can find
the empirically optimal portfolios with gross-exposure constraint
$c$. The empirical risk and actual risk start to diverge when $c
\geq 2$. The empirical risks are overly optimistic, reaching zero
for the case of 500 stocks with one year daily returns.  On the
other hand, the actual risk increases with the gross exposure
parameter $c$ until it reaches its asymptote.  Hence, the Markowitz
portfolio does not have the optimal actual risk.

Our approach has important implications in practical portfolio
selection and allocation. Monitoring and managing a portfolio of
many stocks is not only time consuming but also expensive.
Therefore, it is ideal to pick a reasonable number of assets to
mitigate these two problems. Ideally, we would like to construct a
robust portfolio of reasonably small size to reduce trading,
re-balancing, monitoring, and research costs. We also want to
control the gross exposure of the portfolio to avoid too extreme
long and short positions.  However, to form all optimal subsets of
portfolios of different sizes from a universe of over 2,000 (say)
assets is an NP-hard problem if we use the traditional best subset
approach, which cannot be solved efficiently in feasible time.  Our
algorithm allows one to pick an optimal subset of any number of
assets and optimally allocate them with gross-exposure constraints.
In addition, its associated utility as a function of the number of
selected assets is also available so that the optimal number of
portfolio allocations can be chosen.


\section{Portfolio optimization with gross-exposure constraints}

Suppose we have $p$ assets with returns $R_1$, $\cdots, R_p$ to be
managed. Let $\bR$ be the return vector, $\bSigma$ be its associated
covariance matrix, and $\bw$ be its portfolio allocation vector,
satisfying $\bw^T \bone=1$. Then the variance of the portfolio
return $\bw^T\bR$ is given by $\bw^T \bSigma \bw$.

\subsection{Constraints on gross exposure}

Let $U(\cdot)$ be the utility function, and $\|\bw\|_1 = |w_1| +
\cdots + |w_{p}|$ be the $L_1$ norm. The constraint $\|\bw\|_1 \leq
c$ prevents extreme positions in the portfolio. A typical choice of
$c$ is about 1.6, which results in approximately 130\% long
positions and 30\% short positions\footnote{Let $w^+$ and $w^-$ be
the total percent of long and short positions, respectively. Then,
$w^+ - w^- = 1$ and $w^+ +w^- \leq c$. Therefore, $w^+ \leq (c+1)/2$
and $w^- \leq (c-1)/2$, and $(c-1)/2$ can be interpreted as the
percent of short positions allowed.  Typically, when the portfolio
is optimized, the constraint is usually attained at its boundary
$\|\bw\|_1 = c$.  The constraint on $\|\bw\|_1$ is equivalent to the
constraint on $w^-$. }. When $c=1$, this means that no short sales
are allowed as studied by Jagannathan and Ma (2003). When $c =
\infty$, there is no constraint on short sales. As a generalization
to the work by Markowitz (1952) and Jagannathan and Ma (2003), our
utility optimization problem with gross-exposure constraint is
\begin{eqnarray}
  &\max_{\bw}& E [U(\bw^T \bR) ]     \label{b1} \\
  & s.t.     & \bw^T \mathbf{1}=1\nonumber,
\:\: \|\bw\|_1\leq c\nonumber ,\:\:\bA\bw=\ba.
\end{eqnarray}
The utility function can also be replaced by any risk measures such
as those in Artzner \etal (1999), and in this case the utility
maximization should be risk minimization.

 As to be seen shortly,
the gross-exposure constraint is critical in reducing the
sensitivity of the utility function on the estimation errors of
input vectors such as the expected return and covariance matrix. The
extra constraints $\bA\bw=\ba$ are related to the constraints on
percentage of allocations on each sector or industry. It can also be
the constraint on the expected return of the portfolio.

The $L_1$ norm constraint has other interpretations.  For example,
$\|\bw\|_1$ can be interpreted as the transaction cost.  In this
case, one would subtract the term $\lambda \|\bw\|_1$ from the
expected utility function, resulting in maximizing the modified
utility function
$$
   \max_{\bw}  E [U(\bw^T \bX) ]  - \lambda \| \bw \|_1.
$$
This is equivalent to problem (\ref{b1}).

The question of picking a reasonably small number of assets that
have high utility arises frequently in practice.  This is equivalent
to impose the constraint $\|\bw\|_0 \leq c$, where $\| \bw\|_0$ is
the $L_0$-norm, counting number of non-vanishing elements of $\bw$.
The utility optimization with $L_0$-constraint is an NP-complete
numerical optimization problem.  However, replacing it by the $L_1$
constraint is a feasible convex optimization problem.  Donoho and
Elad (2003) gives the sufficient conditions under which two problems
will yield the same solution.

\subsection{Utility and risk approximations}

It is well known that when the return vector $\bR \sim N(\bmu,
\bSigma)$ and  $U(x) = 1 - \exp(-A x)$, with $A$ being the absolute
risk aversion parameter, maximizing the expected utility is
equivalent to maximizing the Markowitz mean-variance function:
\begin{equation}
   M(\bmu, \bSigma) = \bw^T \bmu - \lambda \bw^T \bSigma \bw,
      \label{b2}
\end{equation}
where $\lambda = A/2$. The solution to the Markowitz utility
optimization problem (\ref{b2}) is $\bw^{opt} = c_1 \bSigma^{-1}
\bmu + c_2 \bSigma^{-1} \bone$ with $c_1$ and $c_2$ depending on
$\bmu$ and $\bSigma$ as well. It depends sensitively on the input
vectors $\bmu$ and $\bSigma$, and their accumulated estimation
errors. It can result in extreme positions, which makes it
impractical.

These two problems disappear when the gross-exposure constraint
$\|\bw\|_1 \leq c$ is imposed.  The constraint eliminates the
possibility of extreme positions. The sensitivity of utility
function can easily be bounded as follows:
\begin{equation}
  | M(\hat{\bmu}, \hat{\bSigma}) - M(\bmu, \bSigma)| \leq
  \|\hat{\bmu} - \bmu\|_\infty \|\bw\|_1 +
  \lambda \| \hat{\bSigma} - \bSigma \|_\infty \|\bw\|_1^2,
  \label{b3}
\end{equation}
where $\|\hat{\bmu} - \bmu\|_\infty$ and $\| \hat{\bSigma} - \bSigma
\|_\infty$ are the maximum componentwise estimation error.
Therefore, as long as each element is estimated well, the overall
utility is approximated well without any accumulation of estimation
errors. In other words, even though tens or hundreds of thousands of
parameters in the covariance matrix are estimated with errors, as
long as $\|\bw\|_1 \leq c$ with a moderate value of $c$, the utility
approximation error is controlled by the worst elementwise
estimation error, without any accumulation of errors from other
elements.  The story is very different in the case that there is no
constraint on the short-sale in which $c=\infty$ or more precisely
$c \geq \|\bw^{opt}\|_1$, the $L_1$ norm of Markowitz's optimal
allocation vector. In this case, the estimation error does
accumulate and they are negligible only for a portfolio with a
moderate size, as demonstrated in Fan \etal (2008).

Specifically, if we consider the risk minimization with no
short-sale constraint, then analogously to (\ref{b3}), we have
\begin{equation}
   | R(\bw, \hat{\bSigma}) - R(\bw, \bSigma)| \leq
    \| \hat{\bSigma} - \bSigma \|_\infty \| \bw \|_1^2,
  \label{b4}
\end{equation}
where as in Jagannathan and Ma (2003) the risk is defined by $R(\bw,
\bSigma) = \bw^T \bSigma \bw$.  The most accurate upper bound in
(\ref{b4}) is when $\|\bw\|_1 = 1$, the no-short-sale portfolio, in
this case,
\begin{equation}
   | R(\bw, \hat{\bSigma}) - R(\bw, \bSigma)| \leq
    \| \hat{\bSigma} - \bSigma \|_\infty .
  \label{b5}
\end{equation}

The inequality (\ref{b5}) is the mathematics behind the conclusions
drawn in the seminal paper by Jagannathan and Ma (2003). In
particular, we see easily that estimation errors from (\ref{b5}) do
not accumulate in the risk. Even when the constraint is wrong, we
lose somewhat in terms of theoretical optimal risk, yet we gain
substantially the reduction of the error accumulation of statistical
estimation.  As a result, the actual risks of the optimal portfolios
selected based on wrong constraints from the empirical data can
outperform the Markowitz portfolio.

Note that the results in (\ref{b3}) and (\ref{b4}) hold for any
estimation of covariance matrix.  The estimate $\hat{\bSigma}$ is
not even required to be a semi-definite positive matrix. Each of its
elements is allowed to be estimated separately from a different
method or even a different data set. As long as each element is
estimated precisely, the theoretical minimum risk we want will be
very closed to the risk we get by using empirical data, thanks to
the constraint on the gross exposure. See also Theorems 1--3 below.
This facilitates particularly the covariance matrix estimation for
large portfolios using high-frequency data (Barndorff-Nielsen \etal,
2008) with non-synchronized trading.  The covariance between any
pairs of assets can be estimated separately based on their pair of
high frequency data. For example, the refresh time subsampling in
Barndorff-Nielsen \etal (2008) maintains far more percentage of
high-frequency data for any given pair of stocks than for all the
stocks of a large portfolio. This provides a much better estimator
of pairwise covariance and hence more accurate risk approximations
(\ref{b3}) and (\ref{b4}).  For covariance between illiquid stocks,
one can use low frequency model or even a parametric model such as
GARCH models (see Engle, 1995; Engle \etal, 2008). For example, one
can use daily data along with a method in Engle \etal (2008) to
estimate the covariance matrix for a subset of relatively illiquid
stocks.

Even though we only consider the unweighted constraints on
gross-exposure constraint throughout the paper to facilitate the
presentation, our methods and results can be extended to a weighted
one:
$$
\|\bw\|_a = \sum_{i=1}^p a_i |w_i| \leq c,
$$
for some positive weights $\{a_j\}$ satisfying $\sum_{j=1}^p a_j =
1$. In this case, (\ref{b3}) is more generally bounded by
$$
 | M(\hat{\bmu}, \hat{\bSigma}) - M(\bmu, \bSigma)| \leq
 c \max_{j} |\hat{\mu}_j -
\mu_j|/a_j+  c^2 \max_{i,j} |\hat{\sigma}_{ij} -
\sigma_{ij}|/(a_ja_j) ,
$$
where $\sigma_{ij}$ and $\hat{\sigma}_{ij}$ are the $(i, j)$
elements of $\bSigma$ and $\hat{\bSigma}$, respectively.  The
weights can be used to downplay those stocks whose covariances can
not be accurately estimated, due to the availability of its sample
size or volatility, for example.

\subsection{Risk optimization: some theory}

To avoid the complication of notation and difficulty associated with
estimation of the expected return vector, from now on, we consider
the risk minimization problem (\ref{b5}):
\begin{equation}
    \min_{\bw^T \bsone = 1, \; \; \|\bw\|_1 \leq c } \bw^T \bSigma \bw. \label{b6}
\end{equation}
This is a simple quadratic programming problem\footnote{The
constraint $\|\bw\|_1 \leq c$ can be expressed as $-v_i\leq w_i\leq
v_i, \sum_{i=1}^{p}v_i\leq c$.  Alternatively, it can be expressed
as $\sum_{i=1}^p w_i^+ - \sum_{i=1}^p w_i^- \leq c$ and $w_i^+ \geq
0$ and $w_i^- \geq 0$. Both expressions are linear constraints and
can be solved by a quadratic programming algorithm.} and can be
solved easily numerically for each given $c$. The problem with
sector constraints can be solved similarly by substituting the
constraints into (\ref{b6}) \footnote{For sector or industry
constraints, for a given sector with $N$ stocks, we typically take
an ETF on the sector along with other $N-1$ stocks as $N$ assets in
the sector. Use the sector constraint to express the weight of the
ETF as a function of the weights of $N-1$ stocks. Then, the
constraint disappears and we need only to determines the $N-1$
weights from problem (\ref{b6}).}.

To simplify the notation, we let
\begin{equation}
R(\bw) = \bw^T \bSigma \bw, \qquad R_n(\bw) = \bw^T \hat{\bSigma}
\bw, \label{b7}
\end{equation}
be respectively the theoretical and empirical portfolio risks with
allocation $\bw$, where $\hat{\bSigma}$ is an estimated covariance
matrix based on the data with sample size $n$.  Let
\begin{equation}
 \bw_{opt} = \argmin_{\bw^T \bsone = 1, \; ||\bw||_1 \leq c}
    R(\bw), \qquad  \hat \bw_{opt}= \argmin_{\bw^T \bsone = 1, \; ||\bw||_1 \leq c}
    \mbox{R}_n(\bw)
    \label{b8}
\end{equation}
be respectively the theoretical optimal allocation vector we want
and empirical optimal allocation vector we get.\footnote{The
solutions depend, of course, on $c$ and their dependence is
suppressed.  The solutions $\bw_{opt}(c)$ and  $\hat \bw_{opt}(c)$
as a function of $c$ are called solution paths.} The following
theorem shows the theoretical minimum risk $R(\bw_{opt})$ (also
called the oracle risk) and the actual risk $R(\hat \bw_{opt})$ of
the invested portfolio are approximately the same as long as the $c$
is not too large and the accuracy of estimated covariance matrix is
not too poor. Both of these risks are unknown. The empirical minimum
risk $R_n(\hat{\bw}_{opt})$ is known, and is usually overly
optimistic. But, it is close to both the theoretical risk and the
actual risk when $c$ is moderate (see Figure 1) and the elements in
the covariance matrix is well estimated.  The concept of risk
approximation is similar to persistent in statistics (Greentshein
and Ritov, 2005).

\begin{theorem}  
Let $a_n = \|\hat{\bSigma} - \bSigma\|_\infty$.  Then, we have
\begin{eqnarray*}
   |R(\hat\bw_{opt}) - R(\bw_{opt})| & \leq & 2 a_nc^2\\
   |R(\hat\bw_{opt}) - R_n(\hat{\bw}_{opt})| & \leq & a_nc^2,
\end{eqnarray*}
and
$$
   |R(\bw_{opt}) - R_n(\hat \bw_{opt})| \leq 3 a_nc^2.
$$
\end{theorem}

Theorem 1 gives the upper bounds on the approximation errors, which
depend on the maximum of individual estimation errors in the
estimated covariance matrix.  There is no error accumulation
component in Theorem 1, thanks to the constraint on the gross
exposure. In particular, the no short-sale constraint corresponds to
the specific case with $c = 1$, which is the most conservative case.
The result holds for more general
$c$.  
As noted at the end of \S2.2, the covariance matrix $\hat{\bSigma}$
is not required to be semi-positive definite, and each element can
be estimated by a different method or data sets, even without any
coordination.  For example, some elements such as the covariance of
illiquid assets can be estimated by parametric models and other
elements can be estimated by using nonparametric methods with
high-frequency data. One can estimate the covariance between $R_i$
and $R_j$ by simply using
\begin{equation}
      \cov(R_i, R_j) = [\var(R_i + R_j) - \var(R_i-R_j)]/4,
      \label{b9}
\end{equation}
as long as we know how to estimate univariate volatilities of the
portfolios $\{R_i+R_j\}$ and$\{R_i - R_j\}$ based on high-frequency
data. While the sample version of the estimates (\ref{b9}) might not
form a semi-positive definite covariance matrix, Theorem 1 is still
applicable.  This allows one to even apply univariate GARCH models
to estimate the covariance matrix, without facing the curse of
dimensionality.

In Theorem 1, we do not specify the rate $a_n$.  This depends on the
model assumption and method of estimation.  For example, one can use
the factor model to estimate the covariance matrix as in Jagannathan
and Ma (2003), Ledoit and Wolf (2004), and Fan \etal
(2008).\footnote{The factor model with known factors is the same as
the multiple regression problem (Fan \etal 2008). The regression
coefficients can be estimated with root-$n$ consistent. This
model-based estimator will not give a better rate of convergence in
terms of $a_n$ than the sample covariance matrix, but with a smaller
constant factor.  When the factor loadings are assumed to be the
same, the rate of convergence can be improved.} One can also
estimate the covariance via the dynamic equi-correlation model of
Engle and Kelly (2007) or more generally dynamically equi-factor
loading models.  One can also aggregate the large covariance matrix
estimation based on the high frequency data (Andersen \etal, 2003,
Barndorff-Nielsen and Shephard, 2002; A\"it-Sahalia, \etal, 2005;
Zhang, \etal, 2005; Patton, 2008) and some components based on
parametric models such as GARCH models. Different methods have
different model assumptions and give different accuracies.

To understand the impact of the portfolio size $p$ on the accuracy
$a_n$, let us consider the sample covariance matrix $\bS_n$ based on
a sample $\{\bR_t\}_{t=1}^n$ over $n$ periods.  This also gives
insightful explanation why risk minimization using sample covariance
works for large portfolio when the constraint on the gross exposure
is in place (Jagannathan and Ma, 2003).  We assume herewith that $p$
is large relative to sample size to reflect the size of the
portfolio, i.e., $p = p_n \to \infty$. When $p$ is fixed, the
results hold trivially.

\begin{theorem}   
Under Condition 1 in the Appendix, we have
$$
        \|\hat{\bS}_n - \bSigma \|_{\infty} = O_p(\sqrt{\frac{\log p}{n}}).
$$
\end{theorem}

This theorem shows that the portfolio size enters into the maximum
estimation error only at the logarithmic order.  Hence, the
portfolio size does not play a significant role in risk minimization
as long as the constraint on gross exposure is in place.  Without
such a constraint, the above conclusion is in general false.

In general, the uniform convergence result typically holds as long
as the estimator of each element of the covariance matrix is
root-$n$ consistent with exponential tails.

\begin{theorem}  
Let $\sigma_{ij}$ and $\hat{\sigma}_{ij}$ be the $(i,j)$th element
of the matrices $\bSigma$ and $\hat{\bSigma}$, respectively.  If for
a sufficiently large $x$,
$$
   \max_{i, j} P\{ \sqrt{n} | \sigma_{ij} - \hat{\sigma}_{ij}| > x \} <
      \exp( - C x^{1/a}),
$$
for two positive constants $a$ and $C$, then
\begin{equation}
\|\bSigma - \hat{\bSigma} \|_\infty = O_P \left ( \frac{ (\log
p)^{a}}{\sqrt{n}} \right ).   \label{b10}
\end{equation}
In addition, if Condition 2 in Appendix holds, then (\ref{b10})
holds for sample covariance matrix, and if Condition 3 holds, then
(\ref{b10}) holds for $a=1/2$.
\end{theorem}

\section{Portfolio tracking and asset selection}

The risk minimization problem (\ref{b6}) has important applications
in portfolio tracking and asset selection.  It also allows one to
improve the utility of existing portfolios.  We first illustrate its
connection to a penalized least-squares problem, upon which the
whole solution path can easily be found (Efron, \etal 2004) and then
outline its applications in finance.

\subsection{Connection with regression problem}

Markowitz's risk minimization problem can be recast as a regression
problem.  By using the fact that the sum of total weights is one, we
have
\begin{eqnarray}
\var(\bw^T\bR) & = & \min_b E(\bw^T\bR - b)^2 \nonumber \\
               & = & \min_b E(Y - w_1 X_1 - \cdots - w_{p-1}
               X_{p-1} - b)^2, \label{c1}
\end{eqnarray}
where $Y = R_p$ and $X_j = R_p - R_j\; (j=1, \cdots, p-1)$. Finding
the optimal weight $\bw$ is equivalent to finding the regression
coefficient $\bw^* = (w_1, \cdots, w_{p-1})^T$ along with the
intercept $b$ to best predict $Y$.

Now, the gross-exposure constraint $\|\bw\|_1 \leq c$ can now be
expressed as $\|\bw^*\|_1 \leq c - | 1 - \bone^T \bw^*|$.  The
latter can not be expressed as
\begin{equation}
   \|\bw^*\|_1 \leq d,   \label{c2}
\end{equation}
for a given constant $d$.  Thus, problem (\ref{b6}) is similar to
\begin{equation}
    \min_{b, \|\bw^*\|_1 \leq d} E(Y - \bw^*{}^T \bX - b)^2,
    \label{c3}
\end{equation}
where $\bX = (X_1, \cdots, X_{p-1})^T$.  But, they are not
equivalent. The latter depends on the choice of asset $Y$, while the
former does not.

Recently, Efron \etal. (2004) developed an efficient algorithm by
using the least-angle regression (LARS), called the LARS-LASSO
algorithm (see Appendix B), to efficiently find the whole solution
path $\bw_{opt}^*(d)$, for all $d \geq 0$, to the constrained
least-squares problem (\ref{c3}). The number of non-vanishing
weights varies as $d$ ranges from 0 to $\infty$.  It recruits
successively one stock, two stocks, and gradually all stocks.  When
all stocks are recruited, the problem is the same as the Markowitz
risk minimization problem, since no gross-exposure constraint is
imposed when $d$ is large enough.

\subsection{Portfolio tracking and asset selection}

Problem (\ref{c3}) depends on the choice of the portfolio $Y$.  If
the variable $Y$ is the portfolio to be tracked, problem (\ref{c3})
can be interpreted as finding a limited number of stocks with a
gross-exposure constraint to minimize the expected tracking error.
As $d$ relaxes, the number of selected stocks increases, the
tracking error decreases, but the short percentage increases. With
the LARS-LASSO algorithm, we can plot the expected tracking error
and the number of selected stocks, against $d$. See, for example,
Figure 2 below for an illustration. This enables us to make an
optimal decision on how many stocks to pick to trade off between the
expected tracking errors, the number of selected stocks and short
positions.

Problem (\ref{c3}) can also be regarded as picking some stocks to
improve the performance of an index or an ETF or a portfolio under
tracking.  As $d$ increases, the risk (\ref{c3}) of the
portfolio\footnote{The exposition implicitly assumes here that the
index or portfolio under tracking consists of all $p$ stock returns
$R_1, \cdots, R_p$, but this assumption is not necessary. Problem
(\ref{c3}) is to modify some of the weights to improve the
performance of the index or portfolio. If the index or portfolio is
efficient, then the risk minimizes at $d=0$.}, consisting of
$\bw_{opt}^*(d)$ (most of components are zero when $d$ is small)
allocated on the first $p-1$ stocks and the rest on $Y = X_p$,
decreases and one can pick a small $d_0$ such that the risk fails to
decrease dramatically. Let $\bw_{o}^*$ be the solution to such a
choice of $d_0$ or any value smaller than this threshold to be more
conservative. Then, our selected portfolio is simply to allocate
$\bw_{o}^*$ on the first $p-1$ stocks $R_1, \cdots,R_{p-1}$ and
remaining percentage on the portfolio $R_p$ to be tracked.  If
$\bw_{o}^*$ has 50 non-vanishing coefficients, say, then we
essentially modify 50 weights of the portfolio $Y = R_p$ to be
tracked to improve its performance. Efficient indices or portfolios
correspond to the optimal solution $d_0 = 0$. This also provides a
method to test whether a portfolio under consideration is efficient
or not.

\begin{figure}[tp] \begin{center} 
 \includegraphics[scale=0.5]{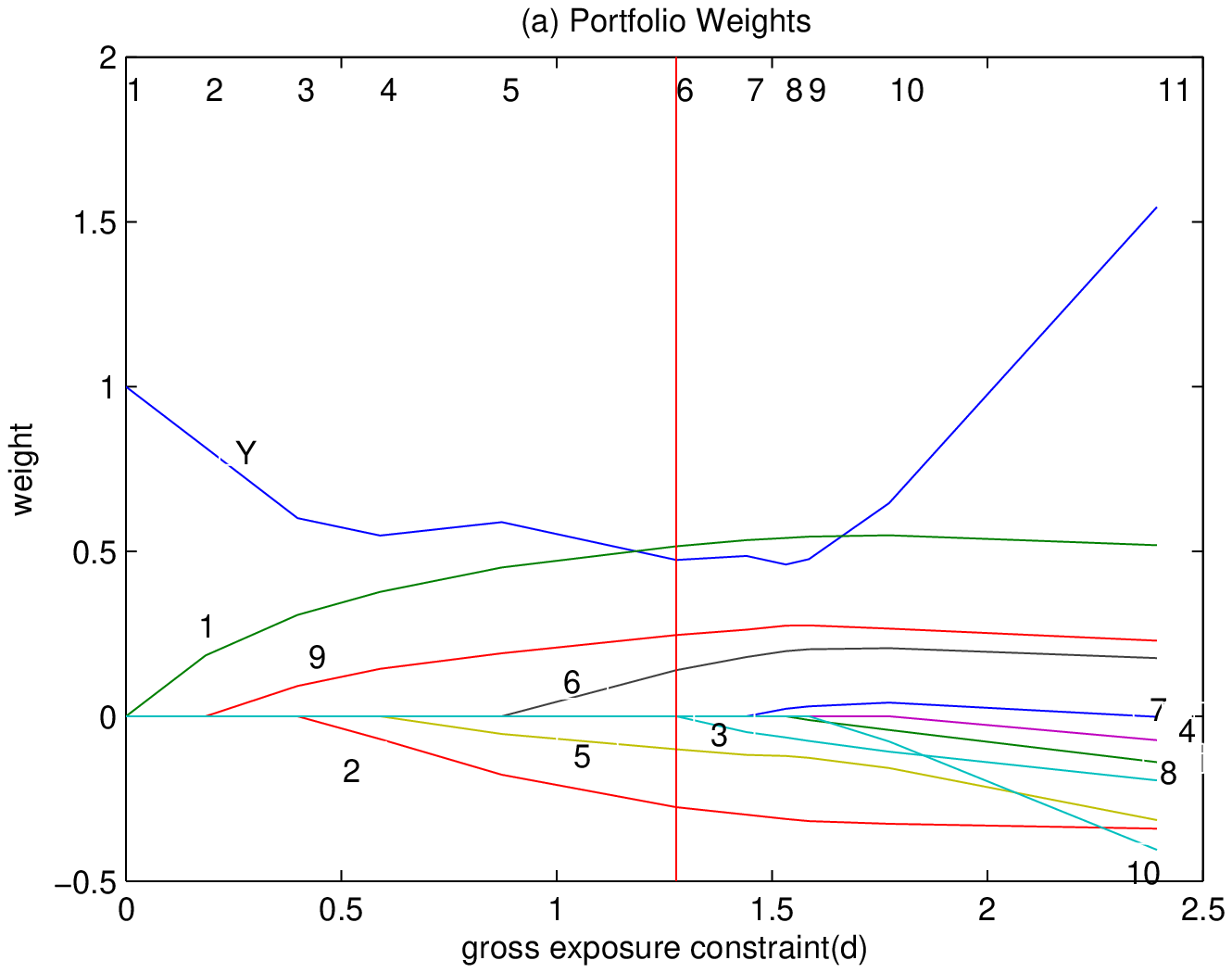}
  \includegraphics[scale=0.5]{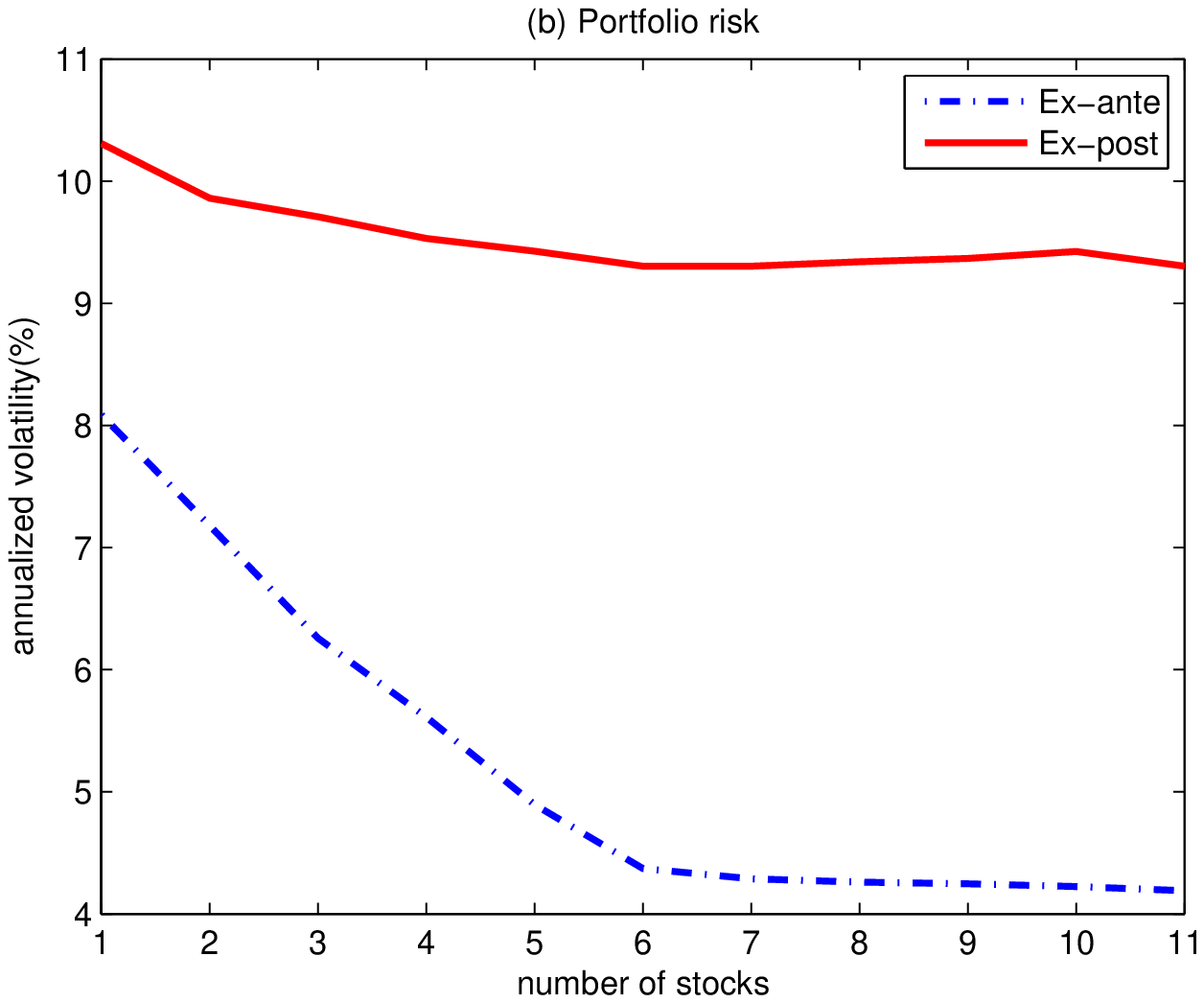}
 \begin{singlespace}
   \caption{\small Illustration of risk improvement by using the
constrained least-squares (\ref{c3}).  On January 8, 2005, it is
intended to improve the risk of the CRSP index using 10 industrial
portfolios constructed by Kenneth French, using the past year daily
data. (a) The solution paths for different gross exposure parameter
$d$, using sample covariance matrix.  The numbers on the top of the
figure shows the number of assets recruited for a given $d$.  (b)
The ex-ante and ex-post risks (annualized volatility) of the
selected portfolio.   Ex-post risk is computed based on the daily
returns of the selected portfolio from January 8, 2005 to January 8,
2006. They have the same decreasing pattern until 6 stocks are
added. }
\end{singlespace}
\end{center}
\end{figure}

As an illustration of the portfolio improvement, we use the daily
returns of 10 industrial portfolios from the website of Kenneth
French from July 1, 1963 to December 31, 2007.  These portfolios are
``Consumer Non-durables", ``Consumer Durables'', ``Manufacturing'',
``Energy'', ``Hi-tech equipment'', ``Telecommunication'', ``Shops'',
``Health'', ``Utilities'', and ``Others''.  They are labeled,
respectively, as 1 through 10 in Figure 2(a).  Suppose that today is
January 8, 2005, which was picked at random, and the portfolio to be
improved is 
the CRSP value-weighted index. We wish to add some of these 10
industrial portfolios to reduce the risk of the index. Based on the
sample covariance matrix, computed from the daily data between
January 9, 2004 and January 8, 2005, we solve problem (\ref{c3})
based on the LARS-LASSO algorithm. The solution path
$\bw^*_{opt}(d)$ is shown in Figure 2(a).  For each given $d$, the
non-vanishing weights of 10 industrial portfolios are plotted along
with the weight on the CRSP.  They add up to one for each given $d$.
For example, when $d=0$, the weight on CRSP is 1.  As soon as $d$
moves slightly away from zero, the ``Consumer Non-durables" (labeled
as 1) are added to the portfolio, while the weight on CRSP is
reduced by the same amount until at the point $d=0.23$, at which the
portfolio ``Utilities'' (labeled as 9) is recruited. At any given
$d$, the weights add up to one. Figure 2(b) gives the empirical
(ex-ante) risk of the portfolio with the allocation vector
$\bw^*_{opt}(d)$ on the 10 industrial portfolios and the rest on the
index. This is available for us at the time to make a decision on
whether or not to modify the portfolio weights.  The figure suggests
that the empirical risk stops decreasing significantly as soon as
the number of assets is more than 6, corresponding to $d = 1.3$,
shown as the vertical line in Figure 2(a).  In other words, we would
expect that the portfolio risk can be improved by adding selected
industrial portfolios until that point. The ex-post risks based on
daily returns until January 8, 2006 (one year) for these selected
portfolios are also shown in Figure 2(b). As expected, the ex-post
risks are much higher than the ex-ante risks. A nice surprise is
that the ex-post risks also decrease until the number of selected
portfolio is 6, which is in line with our decision based on the
ex-ante risks. Investors can make a sensible investment decision
based on the portfolio weights in Figure 2(a) and the empirical
risks in Figure 2(b).

The gaps between the ex-ante and ex-post risks widen as $d$
increases.  This is expected as Theorem 1 shows that the difference
increases in the order of $c^2$, which is related to $d$ by
(\ref{c4}) below.  In particular, it shows that the Markowitz
portfolio has the widest gap.

\subsection{Approximate solution paths to risk minimization}

The solution path to (\ref{c3}) also provides a nearly optimal
solution path to the problem (\ref{b6}).  For example, the
allocation with $\bw_{opt}^*(d)$ on the first $p-1$ stocks and the
rest on the last stock is a feasible allocation vector to the
problem (\ref{b6}) with
\begin{equation}
    c = d + | 1 - \bone^T \bw_{opt}^*(d)|. \label{c4}
\end{equation}
This will not be the optimal solution to the problem (\ref{b6}) as
it depends on the choice of $Y$.  However, when $Y$ is properly
chosen, the solution is nearly optimal, as to be demonstrated. For
example, by taking $Y$ to be the no-short-sale portfolio, then
problem (\ref{c3}) with $d=0$ is the same as the solution to problem
(\ref{b6}) with $c=1$.   We can then use (\ref{c3}) to provide a
nearly optimal solution\footnote{As $d$ increases, so does $c$ in
(\ref{c4}).  If there are multiple $d$'s give the same $c$, we
choose the one having the smaller empirical risk.} to the
gross-exposure constrained risk optimization problem with $c$ given
by (\ref{c4}).

In summary, to compute (\ref{b6}) for all $c$, we first find the
solution with $c=1$ using a quadratic programming.  This yields the
optimal no-short-sale portfolio.  We then take this portfolio as $Y$
in problem (\ref{c3}) and apply the LARS-LASSO algorithm to obtain
the solution path $\bw_{opt}^*(d)$.  Finally, use (\ref{c4}) to
convert $d$ into $c$, namely, regard the portfolio with
$\bw_{opt}^*(d)$ on the first $p-1$ stocks and the rest on the
optimal no-short-sale portfolio as an approximate solution to the
problem (\ref{b6}) with $c$ given by (\ref{c4}). This yields the
whole solution path to the problem (\ref{b6}).  As shown in Figure
3(a) below and the empirical studies, the approximation is indeed
quite accurate.

In the above algorithm, one can also take a tradable index or an ETF
in the set of stocks under consideration as the $Y$ variable and
applies the same technique to obtain a nearly optimal solution. We
have experimented this and obtained good approximations, too.

\subsection{Empirical risk minimization}

First of all, the constrained risk minimization problem (\ref{c3})
depends only on the covariance matrix.  If the covariance matrix is
given, then the solution can be found through the LARS-LASSO
algorithm in Appendix B. However, if the empirical data $\{(\bX_t,
Y_t)_{t=1}^n\}$ are given, one naturally minimizes its empirical
counterpart:
\begin{eqnarray}
\min_{b, \|\bw^*\|_1 \leq d} \sum_{t=1}^n (Y_t - \bw^*{}^T \bX_t -
b)^2. \label{c5}
\end{eqnarray}
Note that by using the connections in \S3.1, the constrained
least-squares problem (\ref{c5}) is equivalent to problem (\ref{c3})
with the population covariance matrix replaced by the sample
covariance matrix: No details of the original data are needed and
the LARS-LASSO algorithm in Appendix B applies.

\section{Simulation studies}

In this section, we use simulation studies, in which we know the
true covariance matrix and hence the actual and theoretical risks,
to verify our theoretical results and to quantify the finite sample
behaviors. In particular, we would like to demonstrate that the risk
profile of the optimal no-short-sale portfolio can be improved
substantially and that the LARS algorithm yields a good approximate
solution to the risk minimization with gross-exposure constraint. In
addition, we would like to demonstrate that when covariance matrix
is estimated with reasonable accuracy, the risk that we want and the
risk that we get are approximately the same for a wide range of the
exposure coefficient. When the sample covariance matrix is used,
however, the risk that we get can be very different from the risk
that we want for the unconstrained Markowitz mean-variance
portfolio.

Throughout this paper,  the risk is referring to the standard
deviation of a portfolio, the square-roots of the quantities
presented in Theorem 1.  To avoid ambiguity, we call
$\sqrt{R(\bw_{opt})}$ the theoretical optimal risk or oracle risk,
$\sqrt{R_n(\hat{\bw}_{opt})}$ the empirical optimal risk, and
$\sqrt{R(\hat{\bw}_{opt})}$ the actual risk of the empirically
optimally allocated portfolio. They are also referred to as the
oracle, empirical, and actual risks.

\subsection{A simulated Fama-French three-factor model}

Let $R_i$ be the excessive return over the risk free interest rate.
Fama and French (1993) identified three key factors that capture the
cross-sectional risk in the US equity market.  The first factor is
the excess return of the proxy of the market portfolio, which is the
value-weighted return on all NYSE, AMEX and NASDAQ stocks (from
CRSP) minus the one-month Treasury bill rate. The other two factors
are constructed using six value-weighted portfolios formed by size
and book-to-market. They are the difference of returns between large
and small capitalization, which captures the size effect, and the
difference of returns between high and low book-to-market ratios,
which reflects the valuation effect.  More specifically,  we assume
that the excess return follows the following three-factor model:
\begin{equation} \label{d1}
    R_{i}=b_{i1}f_1+b_{i2}f_2+b_{i3}f_3+ \veps_i, \quad
    i=1,\cdots,p,
\end{equation}
where $\{b_{ij}\}$ are the factor loadings of the $i^{th}$ stock on
the factor $f_j$, and $\varepsilon_i$ is the idiosyncratic noise,
independent of the three factors.  We assume further that the
idiosyncratic noises are independent of each other, whose marginal
distributions are the Student-$t$ with degree of freedom 6 and
standard deviation $\sigma_i$.

To facilitate the presentation, we write the factor model (\ref{d1})
in the matrix form:
\begin{equation} \label{d2}
  \bR = \bB \bff + \bveps,
\end{equation}
where $\bB$ is the matrix, consisting of the factor loading
coefficients. Throughout this simulation, we assume that
$E(\bveps|\bff)=\bzero$ and $\cov(\bveps|\bff)= \diag(\sigma_1^2,
\cdots, \sigma_p^2)$. Then, the covariance matrix of the factor
model is given by
\begin{equation} \label{d3}
    \Sig=\cov(\bB\bff)+\cov(\bveps)=\bB\cov(\bff)\bB^T
          +\mbox{diag}(\sigma_1^2, \cdots, \sigma_p^2).
\end{equation}

We simulate the $n$-period returns of $p$ stocks as follows. See Fan
\etal (2008) for additional details.  First of all, the factor
loadings are generated from the trivariate normal distribution
$N(\bmu_{b},\mbox{\bf cov}_{b})$, where the parameters are given in
Table 1 below.  Once the factor loadings are generated, they are
taken as the parameters and thus kept fixed throughout simulations.
The levels of idiosyncratic noises are generated from a gamma
distribution with shape parameter 3.3586 and the scale parameter
0.1876, conditioned on the noise level of at least 0.1950. Again,
the realizations are taken as $p$ parameters $\{\sigma_i\}$ and kept
fixed across simulations. The returns of the three factors over $n$
periods are drawn from the trivariate normal distribution
$N(\bmu_{f},\mbox{\bf cov}_{f})$, with the parameters given in Table
1 below. They differ from simulations to simulations and are always
drawn from the trivariate normal distribution. Finally, the
idiosyncratic noises are generated from the Student's t-distribution
with degree of freedom 6 whose standard deviations are equal to the
noise level $\{\sigma_i\}$.  Note that both the factor returns and
idiosyncratic noises change across different time periods and
different simulations.

\begin{singlespace}
\begin{table}[tp]   
\begin{center}
\caption{\bf Parameters used in the simulation}
\end{center}
\vspace*{-0.3in}

\noindent \small This table shows the expected values and covariance
matrices for the factor loadings (left panel) and factor returns
(right panel). They are used to generate factor loading parameters
and the factor returns over different time periods.  They were
calibrated to the market.

\begin{center}
\begin{tabular}
[c]{c|rrr c c|rrr} \hline \multicolumn{4}{c}{Parameters for factor
loadings} & & \multicolumn{4}{c}{Parameters for factor returns}\\
\cline{1-4} \cline{6-9}
 \multicolumn{1}{c}{$\bmu_{b}$}  & \multicolumn{3}{c}{$\mbox{\bf cov}_{b}$}  & \hspace{0.3 in} &
 \multicolumn{1}{c}{$\bmu_{f}$}  &  \multicolumn{3}{c}{ $\mbox{\bf cov}_{f}$ } \\  \cline{1-4} \cline{6-9}
0.7828 &  0.02914 &  0.02387 &  0.010184 && 0.02355 & 1.2507  & -0.0350 & -0.2042 \\
0.5180 &  0.02387 &  0.05395 & -0.006967 && 0.01298 & -0.0350 &  0.3156 & -0.0023 \\
0.4100 &  0.01018 & -0.00696 &  0.086856 && 0.02071 & -0.2042 &
-0.0023 & 0.1930
 \\\hline
\end{tabular}
\end{center}
\end{table}
\end{singlespace}

The parameters used in the simulation model (\ref{b1}) are
calibrated to the market data from May 1, 2002 to August 29, 2005,
which are depicted in Table 1 and taken from Fan \etal (2008) who
followed closely the instructions on the website of  Kenneth French,
using the three-year daily return data of 30 industrial portfolios.
The expected returns and covariance matrix of the three factors are
depicted in Table 1. They fitted the data to the Fama-French model
and obtained 30 factor loadings. These factor loadings have the
sample mean vector $\bmu_{b}$ and sample covariance $\mbox{\bf
cov}_{f}$, which are given in Table 1. The 30 idiosyncratic noise
levels were used to determine the parameters in the Gamma
distribution.

\subsection{LARS approximation and portfolio improvement}

Quadratic programming algorithms to problem (\ref{b6}) is relatively
slow when the whole solution path is needed.  As mentioned in \S3.3,
the LARS algorithm provides an approximate solution to this problem
via (\ref{c4}).  The LARS algorithm is designed to compute the whole
solution path and hence is very fast. The first question is then the
accuracy of the approximation.  As a byproduct, we also demonstrate
that the optimal no-short-sale portfolio is not diversified enough
and can be significantly improved.

To demonstrate this, we took 100 stocks with covariance matrix given
by (\ref{d3}).  For each given $c$ in the interval $[1, 3]$, we
applied a quadratic programming algorithm to solve problem
(\ref{b6}) and obtained its associated minimum portfolio risk. This
is depicted in Figure 3(a). We also employed the LARS algorithm
using the optimal no-short-sale portfolio as $Y$, with $d$ ranging
from 0 to 3.  This yields a solution path along with its associated
portfolio risk path. Through the relation (\ref{c4}), we obtained an
approximate solution to problem (\ref{b6}) and its associated risk
which is also summarized in Figure 4(a). The number of stocks for
the optimal no-short-sale portfolio is 9.  As $c$ increases, the
number of stocks picked by (\ref{b6}) also increases, as
demonstrated in Figure 3(b) and the portfolio gets more diversified.

\begin{figure}[tp]   
\begin{center}
\begin{tabular}{c c}
   \includegraphics[scale=0.5]{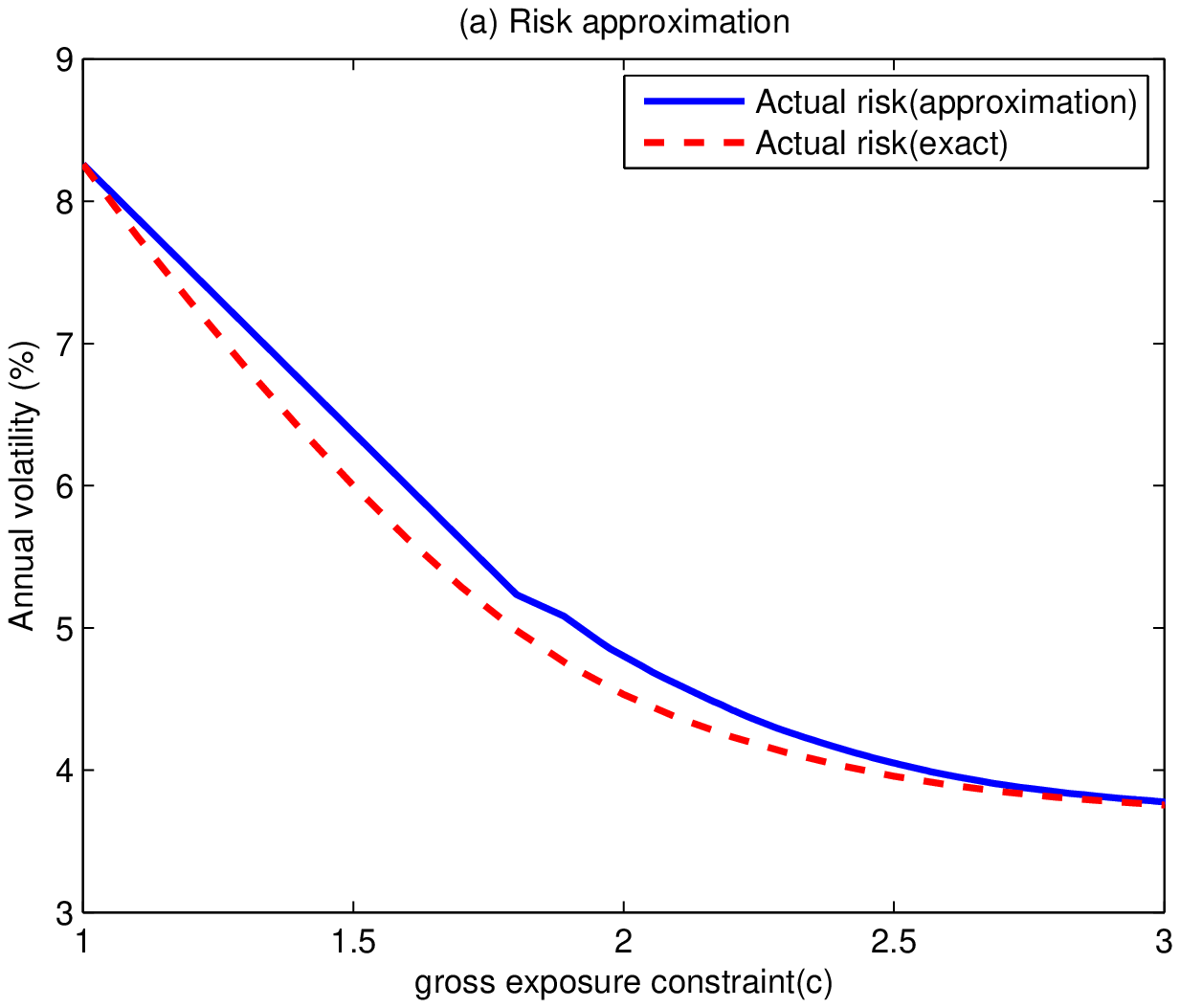} &
    \includegraphics[scale = 0.5]{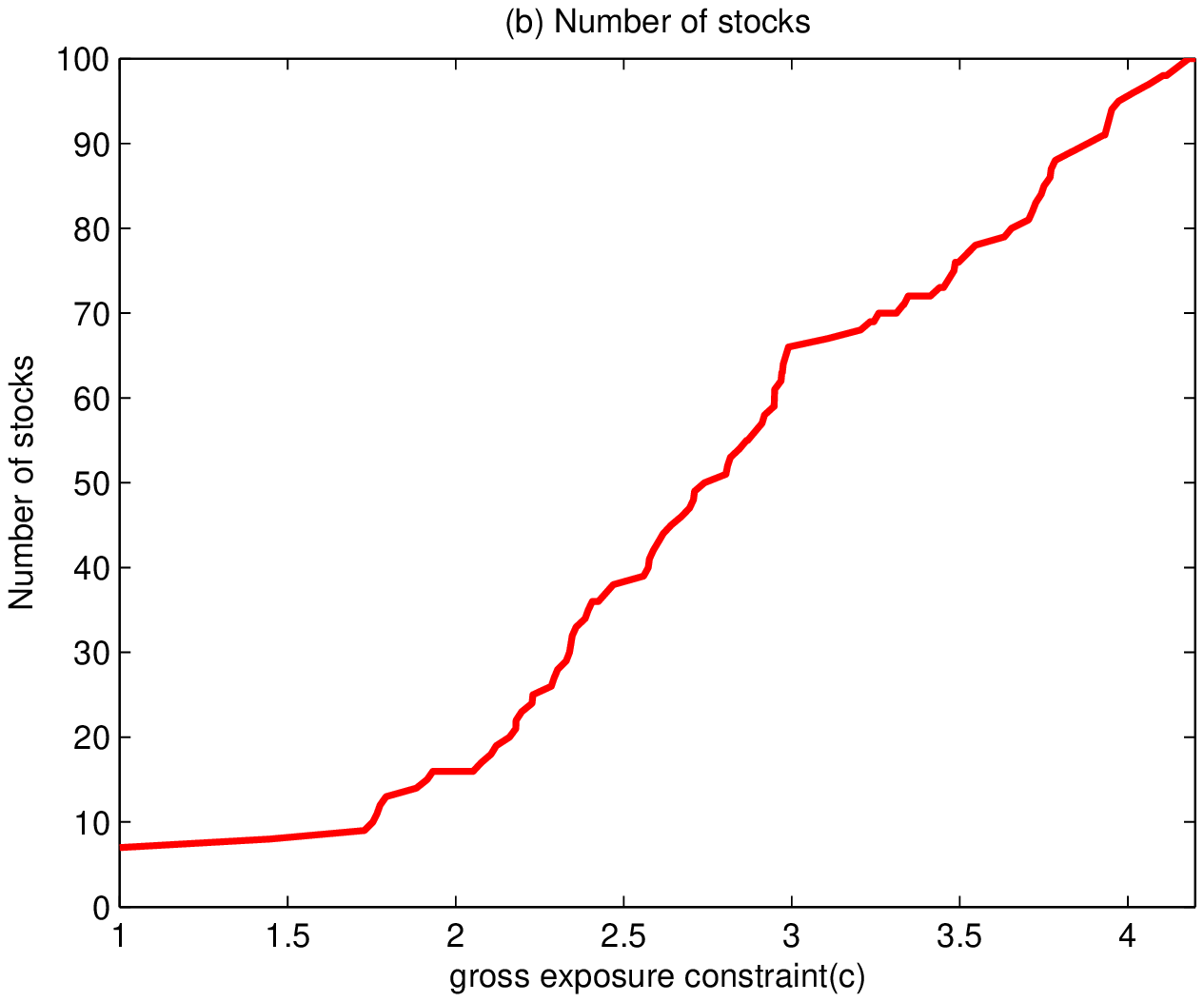} \\
    \includegraphics[scale = 0.5]{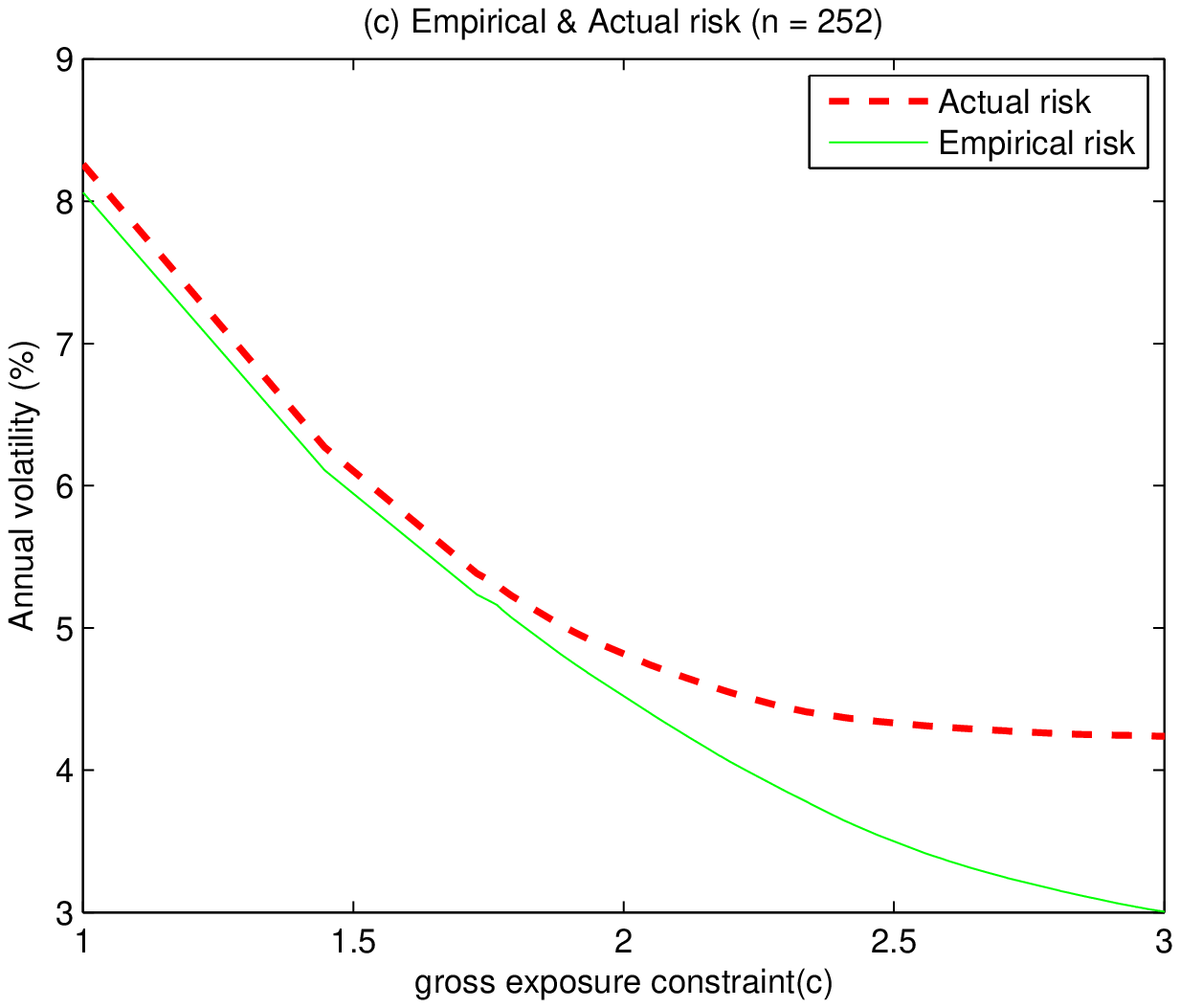} &
    \includegraphics[scale = 0.5]{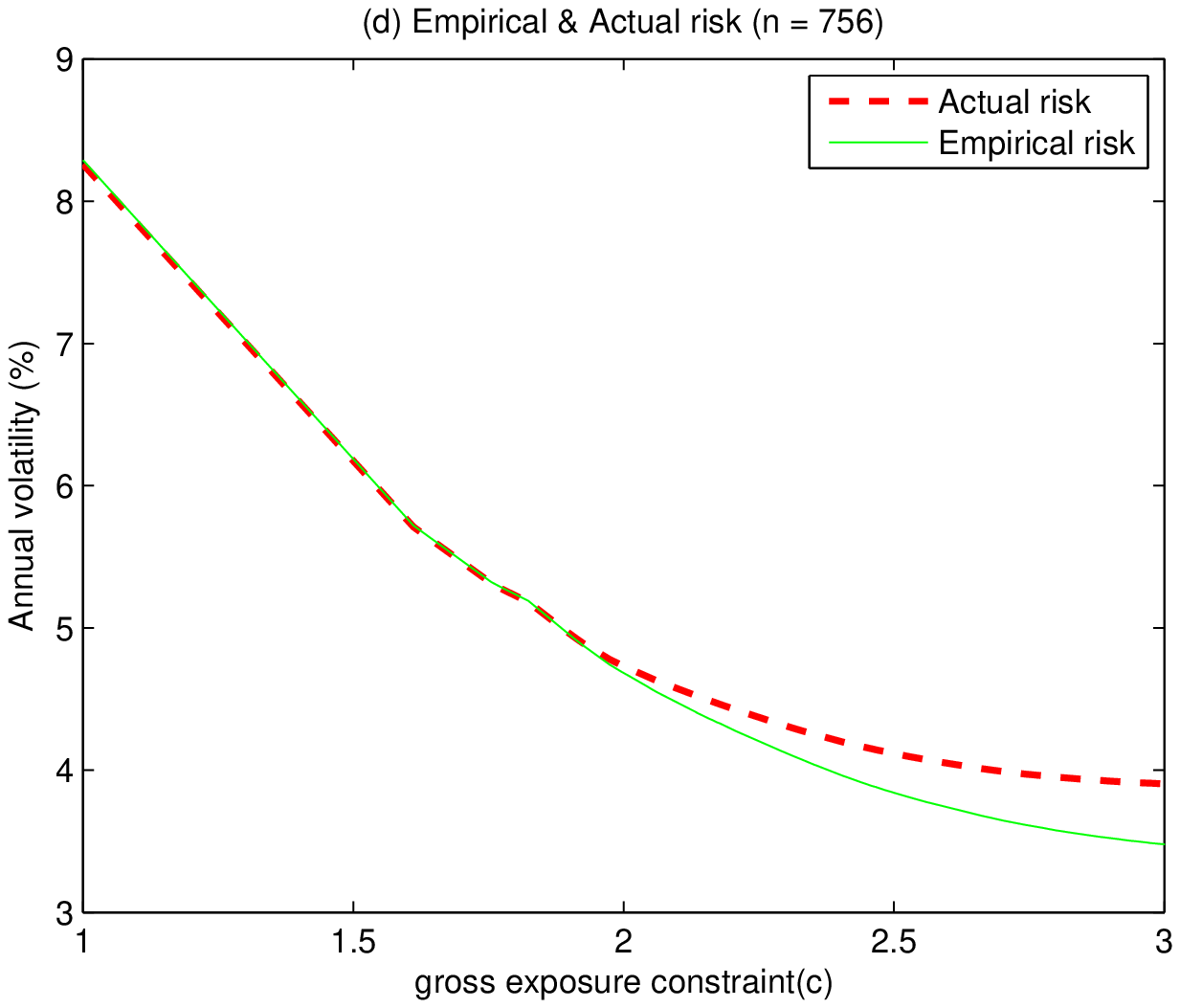}\\
\end{tabular}

\begin{singlespace}
   \caption{\small Comparisons of optimal portfolios selected by
   the exact and approximate algorithms with a known covariance matrix.
     (a) The risks for the exact algorithm
   (dashed line) and the LARS (approximate) algorithm. (b) The number of
    stocks picked by the optimization problem (\ref{b6}) as a function of the
    gross exposure coefficient $c$.  (c)  The actual risk (dashed line)
    and empirical risk (solid) of the portfolio selected
    based on the sample covariance matrix ($n=252$). (d) The same as (c) except
    $n=756$.}
\end{singlespace}
\end{center}
\end{figure}

The approximated and exact solutions have very similar risk
functions.  Figure 3 showed that the optimal no-short-sale portfolio
is very conservative and can be improved dramatically as the
constraint relaxes.  At $c = 2$ (corresponding to 18 stocks with
50\% short positions and 150\% long positions), the risk decreases
from 8.1\% to 4.9\%.  The decrease of risks slows down dramatically
after that point. This shows that the optimal no-short-sale
constraint portfolio can be improved substantially by using our
methods.

The next question is whether the improvement can be realized with
the covariance matrix being estimated from the empirical data.  To
illustrate this, we simulated $n=252$ from the three-factor model
(\ref{d1}) and estimated the covariance matrix by the sample
covariance matrix. The actual and empirical risks of the selected
portfolio for a typical simulated data set are depicted in Figure
3(c). For a range up to $c=1.7$, they are approximately the same.
The range widens when the covariance matrix is estimated with a
better accuracy.  To demonstrate this effect, we show in Figure 3(d)
the case with sample size $n = 756$.  However, when the gross
exposure parameter is large and the portfolio is close to the
Markowitz's one, they differ substantially.   See also Figure 1. The
actual risk is much larger than the empirical one, and even far
larger than the theoretical optimal one. Using the empirical risk as
our decision guide, we can see that the optimal no-short-sale
portfolio can be improved substantially for a range of the gross
exposure parameter $c$.

To demonstrate further how much our method can be used to improve
the existing portfolio, we assume that the current portfolio is an
equally weighted portfolio among 200 stocks.  This is the portfolio
$Y$.  The returns of these 200 stocks are simulated from model
(\ref{d1}) over a period of $n=252$.  The theoretical risk of this
equally weighted portfolio is 13.58\%, while the empirical risk of
this portfolio is 13.50\% for a typical realization.  Here, the
typical sample is referring to the one  that has the median value of
the empirical risks among 200 simulations.   This particular
simulated data set is used for the further analysis.

\begin{figure}[tp]    
\begin{center}
\begin{tabular}{c c}
   \includegraphics[scale=0.5]{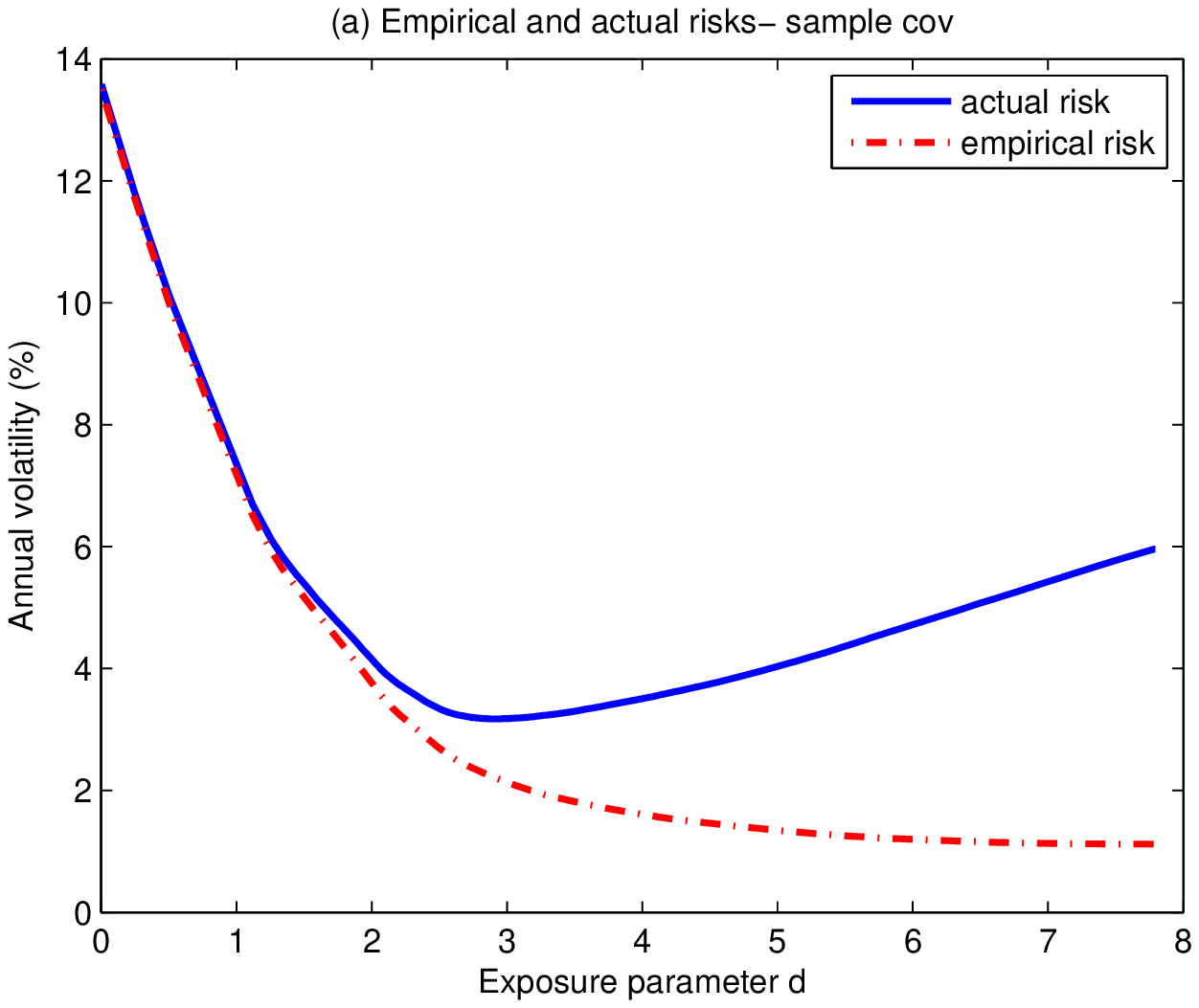} &
    \includegraphics[scale=0.5]{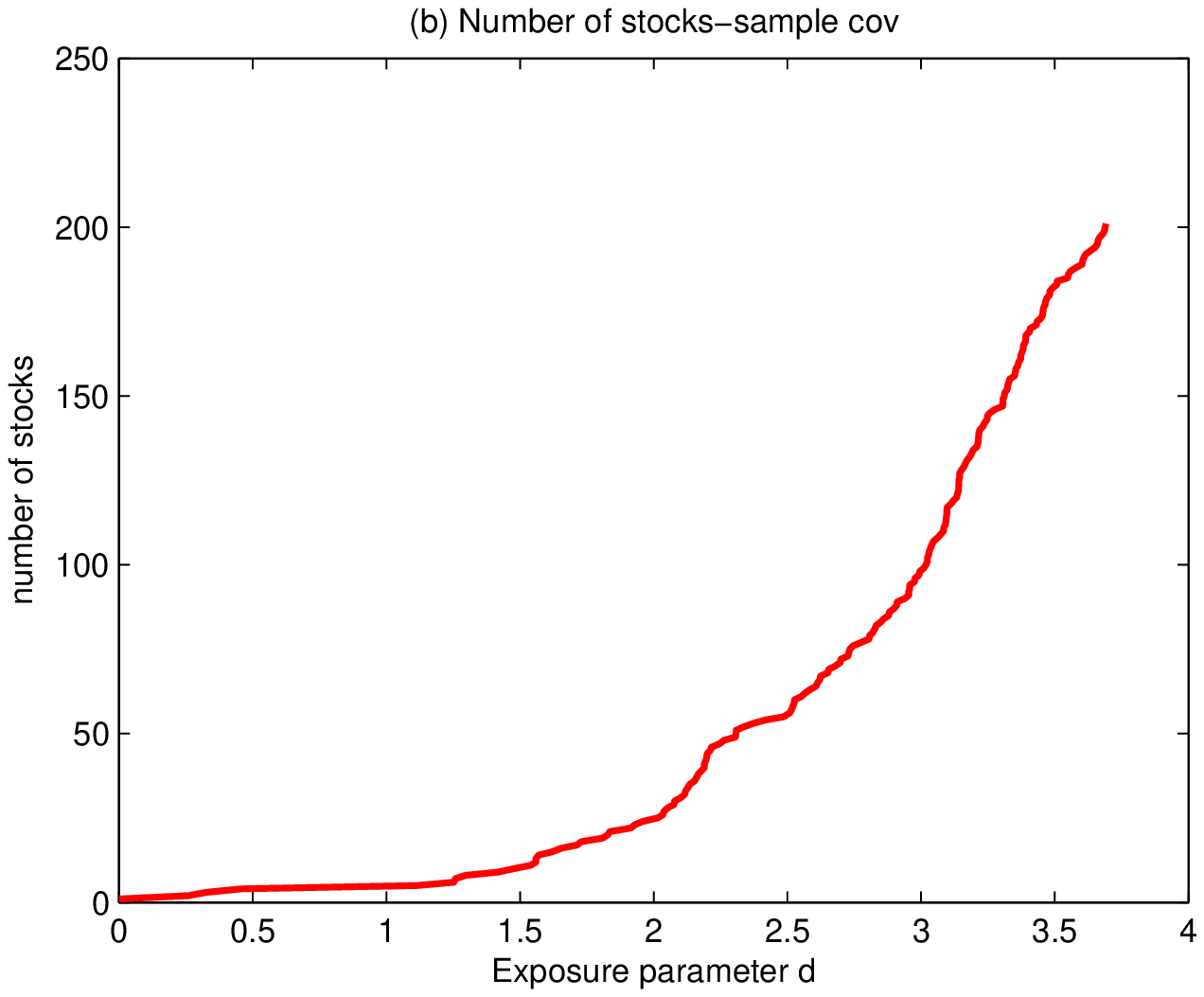}\\
    \includegraphics[scale=0.5]{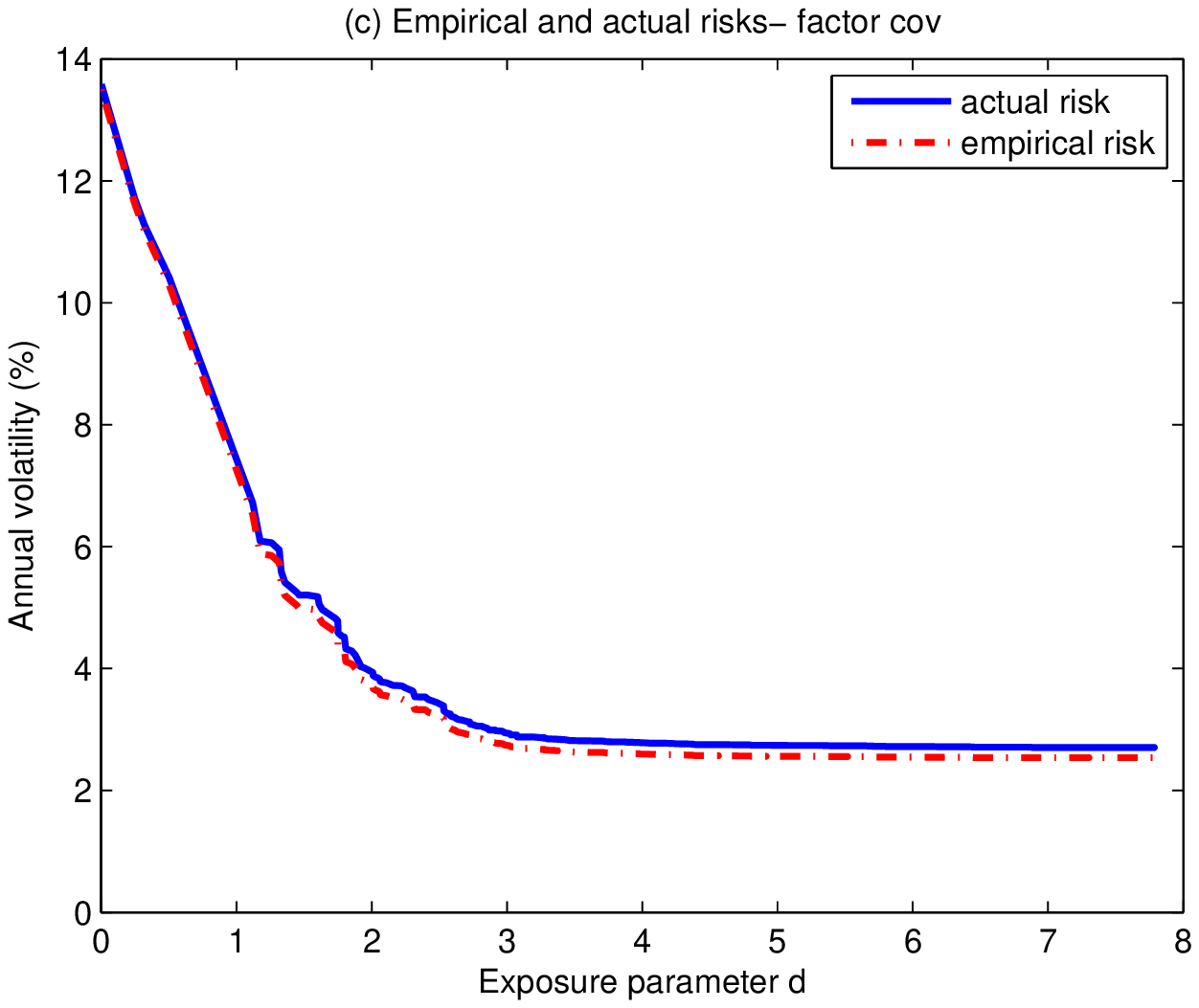} &
    \includegraphics[scale=0.5]{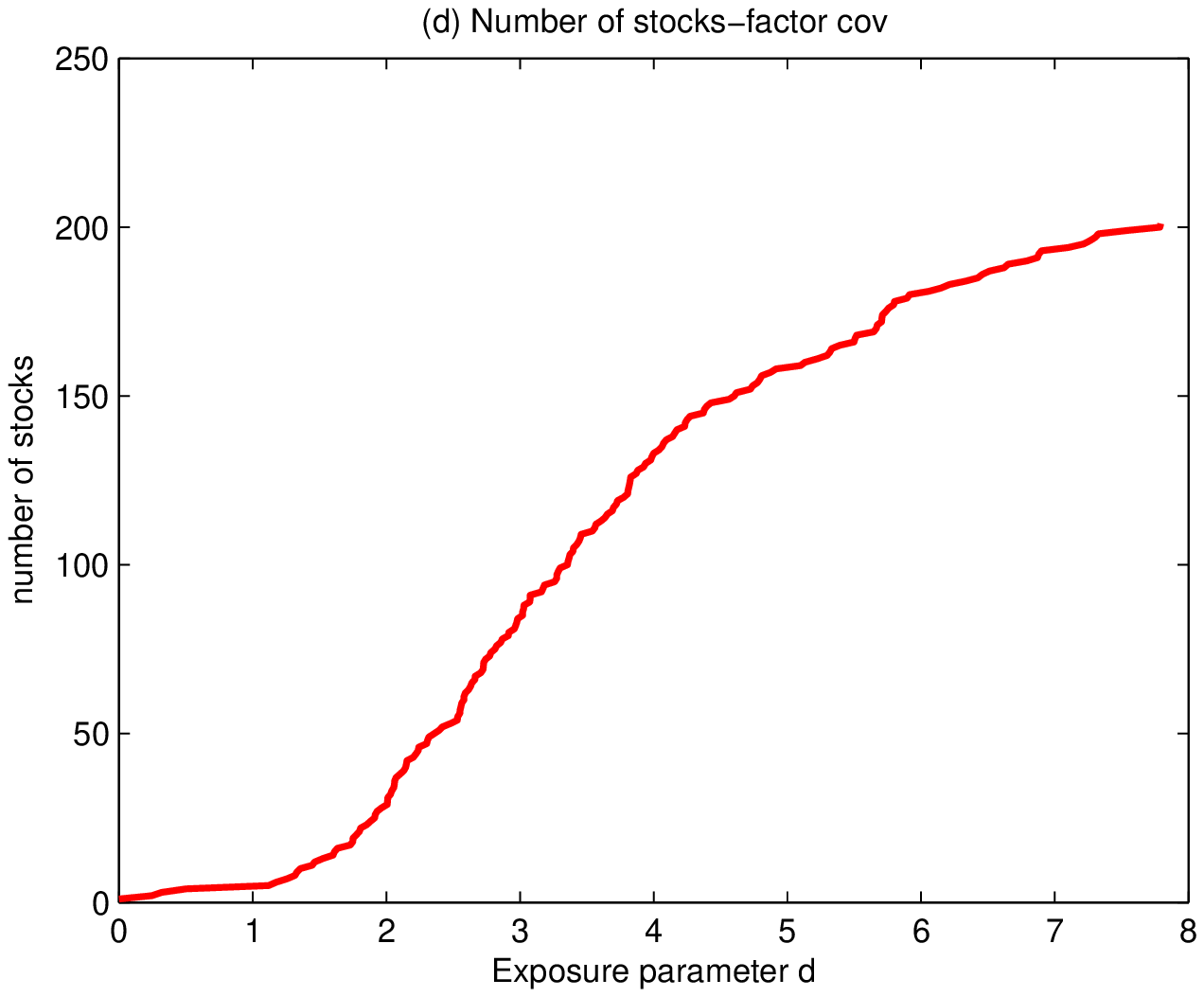}
\end{tabular}

\begin{singlespace}
   \caption{\small This is on the portfolio improvement of the 200
   equally weighted portfolio by modifying the weights of the
   portfolio using (\ref{c3}). As the exposure parameter $d$
   increases, more weights are modified and the risks of new
   portfolios decrease.
   (a)  The empirical and actual risks of the modified portfolios are plotted against exposure parameter $d$, based on the sample covariance
   matrix.
    (b) The number of stocks whose weights are modified as a function of $d$.
    (c) and (d) are the same as (a) and (b) except that the covariance matrix is estimated
    based on the factor model.}
\end{singlespace}
\end{center}
\end{figure}

We now pretend that this equally weighted portfolio is the one that
an investor holds and the investor seeks possible improvement of the
efficiency by modifying some of the weights. The investor employs
the LARS-LASSO technique (\ref{c3}), taking the equally weighted
portfolio as $Y$ and the 200 stocks as potential $X$.  Figure 4
depicts the empirical and actual risks, and the number of stocks
whose weights are modified in order to improve the risk profile of
equally weighted portfolio.

The risk profile of the equally weighted portfolio can be improved
substantially.  When the sample covariance is used,  at $d=1$,
Figure 2(a) reveals the empirical risk is only about one half of the
equally weighted portfolio, while Figure 2(b) or Table 2 shows that
the number of stocks whose weights have been modified is only 4. As
$d = 1$, by (\ref{c4}), $c \leq 2d + 1 \leq 3$, which is a crude
upper bound. In other words, there are at most 100\% short
positions. Indeed, the total percentage of short positions is only
about 48\%. The actual risk of this portfolio is very close to the
empirical one, giving an actual risk reduction of nearly 50\%. At
$d=2$, corresponding to about 130\% of short positions, the
empirical risk is reduced by a factor of about 5, whereas the actual
risk is reduced by a factor of about 4. Increasing the gross
exposure parameter only slightly reduces the empirical risk, but
quickly increases the actual risk. Applying our criterion to the
empirical risk, which is known at the time of decision making, one
would have chosen a gross exposure parameter somewhat less than 1.5,
realizing a sizable risk reduction. Table 2 summarizes the actual
risk, empirical risk, and the number of modified stocks under
different exposure parameter $d$.  Beyond $d = 2$, there is very
little risk reduction.  At $d = 5$, the weights of 158 stocks need
to be modified, resulting in 250\% of short positions.  Yet, the
actual risk is about the same as that with $d=2$.

\begin{singlespace}
\begin{table}[tp]   
\begin{center}
\caption{\bf Empirical and actual risks for selected portfolios}
\end{center}
\vspace*{-0.3 in}

\noindent \small This table is based on a typical simulated 252
daily returns of 200 stocks from the Fama-French three-factor model.
The aim is to improve the risk of the equally weighted portfolio by
modifying some of its weights. The covariance is estimated by either
sample covariance (left panel) or the factor model (right panel).
The penalized least-squares (\ref{c3}) is used to construct the
portfolio. Reported are actual risk, empirical risk, the number of
stocks whose weights are modified by the penalized least square
(\ref{c3}), and percent of short positions, as a function of the
exposure parameter $d$.

\begin{center}
\begin{tabular}
[c]{lccccccccc} \hline
  & \multicolumn{4}{c}{Sample Covariance} &\hspace{0.2 in} &  \multicolumn{4}{c}{Factor-model based covariance} \\
  \cline{2-5} \cline{7-10}
d \hspace*{0.1 in}
  & Actual & Empirical& \# stocks &  Short &
  & Actual & Empirical& \# stocks &  Short \\ \hline
0 & 13.58 & 13.50  & 0    &  0\%  &  & 13.58  & 12.34  & 0    &    0\%  \\
1 & 7.35  & 7.18   & 4    & 48\%  &  & 7.67   & 7.18   & 4    &    78\% \\
2 & 4.27  & 3.86   & 28   & 130\% &  & 4.21   & 4.00   & 2    &    133\%  \\
3 & 3.18  & 2.15   & 84   & 156\% &  & 2.86   & 2.67   & 98   &    151\% \\
4 & 3.50  & 1.61   & 132  & 195\% &  & 2.71   & 2.54   & 200  &    167\%  \\
5 & 3.98  & 1.36   & 158  & 250\% &  & 2.71   & 2.54   & 200  &    167\% \\
\hline
\end{tabular}
\end{center}
\end{table}
\end{singlespace}

Similar conclusions can be made for the covariance matrix based on
the factor model.  In this case, the covariance matrix is estimated
more accurately and hence the empirical and actual risks are closer
for a wider range of the gross exposure parameter $d$.  This is
consistent with our theory. The substantial gain in this case is due
to the fact that the factor model is correct and hence incurs no
modeling biases in estimating covariance matrices.  For the real
financial data, however, the accuracy of the factor model is
unknown. As soon as $d \geq 3$ the empirical reduction of risk is
not significant. The range of risk approximation is wider than that
based on the sample covariance, because the factor-model based
estimation is more accurate.

Figure 4(a) also supports our theory that when $c$ is large, the
estimation errors of covariance matrix start to play a role.  In
particular, when $d=7$, which is close to the Markowitz portfolio,
the difference between actual and empirical risks is substantial.

\subsection{Risk approximations}

We now use simulations to demonstrate the closeness of the risk
approximations with the gross-exposure constraints.  The simulated
factor model (\ref{d1}) is used to generate the returns of $p$
stocks over a period of $n=252$ days.  The number of simulations is
101. The covariance matrix is estimated by either the sample
covariance or the factor model (\ref{d3}) whose coefficients are
estimated from the sample.  We examined two cases:  $p = 200$ and $p
= 500$.  In the first case, the portfolio size is smaller than the
sample size, whereas in the second the portfolio size is larger. The
former corresponds to a non-degenerate sample covariance matrix
whereas the latter corresponds to a degenerate one. The LARS
algorithm is used to find an approximately optimal solution to
problem (\ref{b6}) as it is much faster for the simulation purpose.
We take $Y$ as the optimal portfolio with no-short-sale constraint.

\begin{figure}[t]    
\begin{center}
\begin{tabular}{c c}
   \includegraphics[scale=0.5]{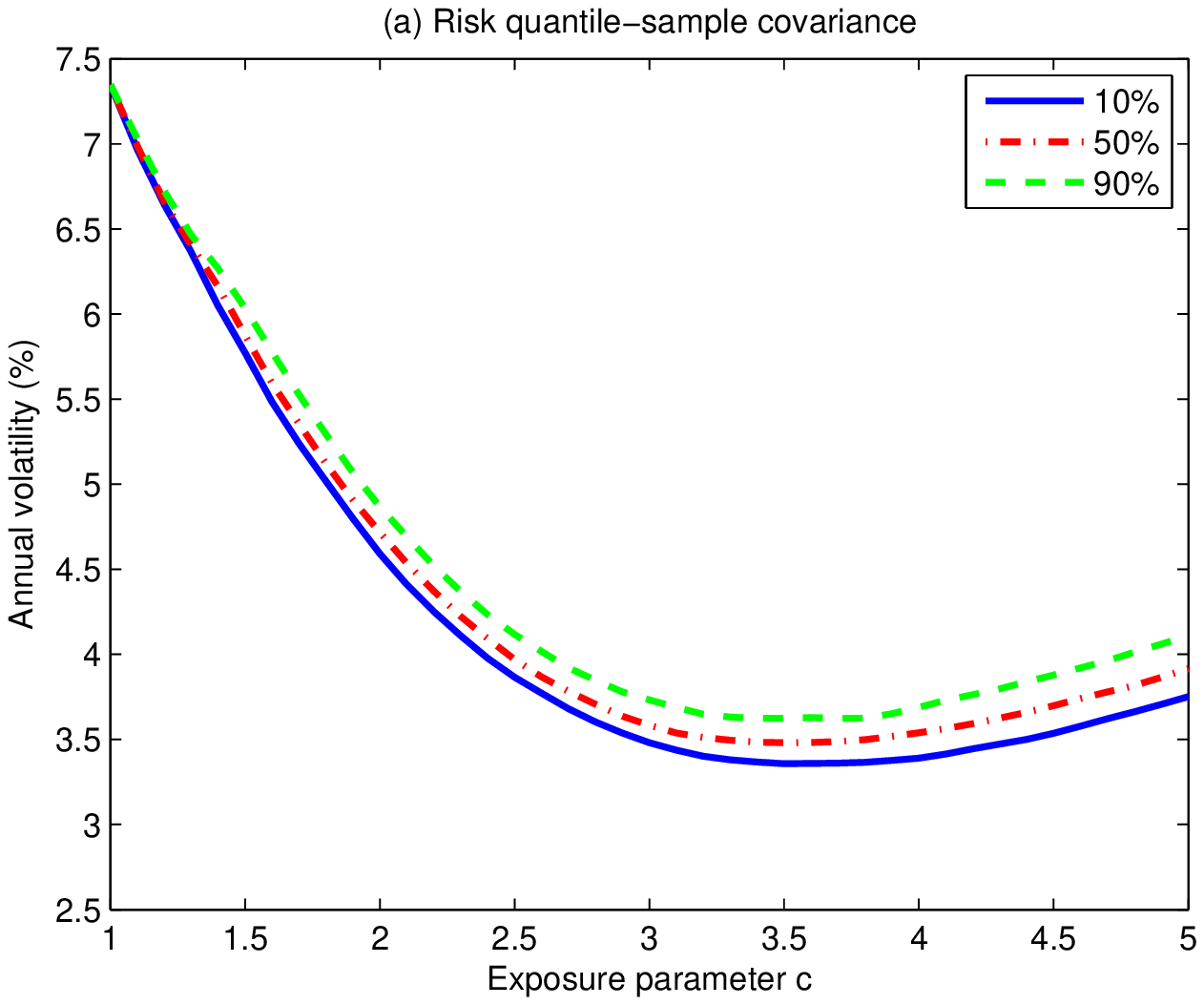} &
   \includegraphics[scale=0.5]{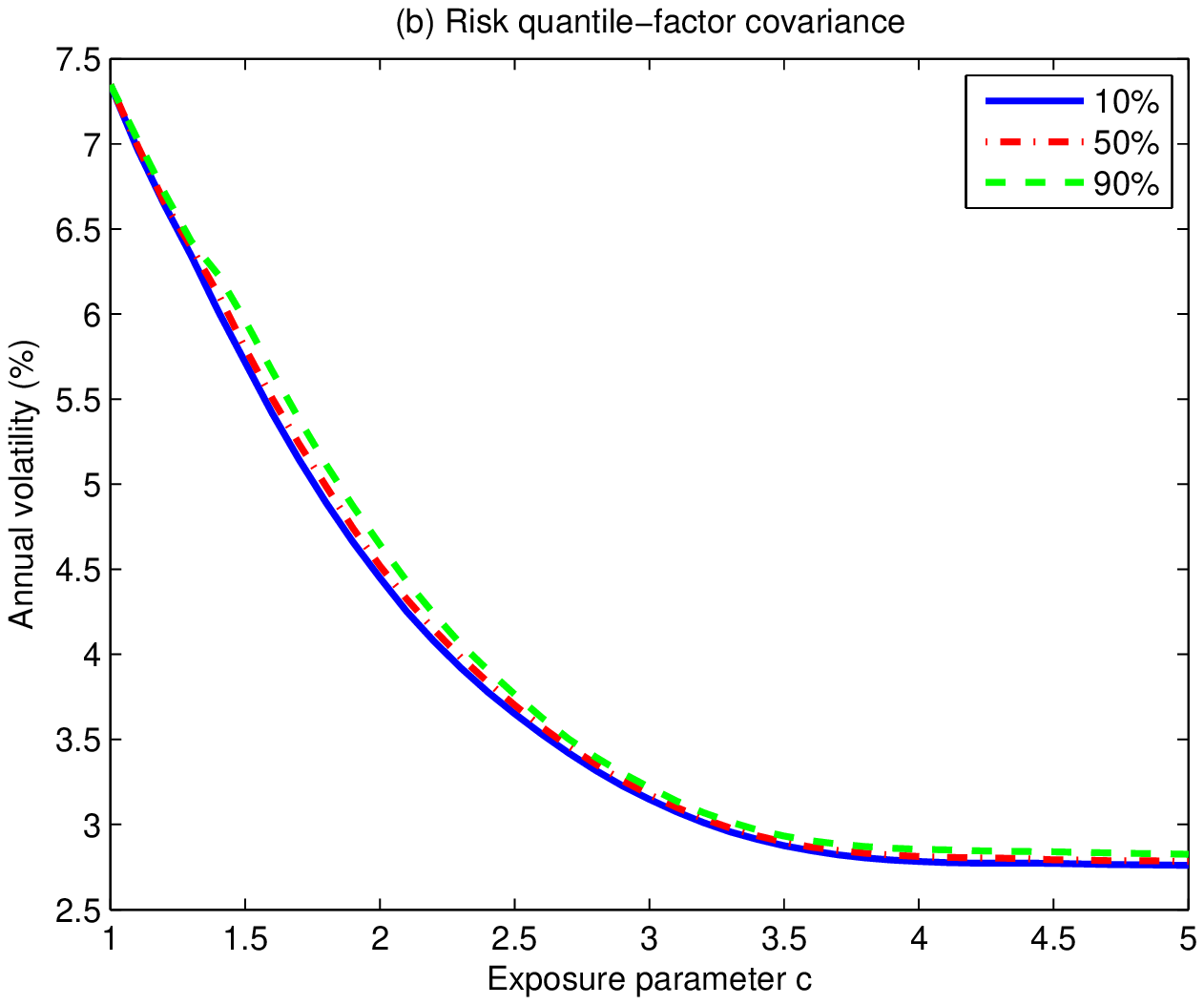} \\
   \includegraphics[scale=0.5]{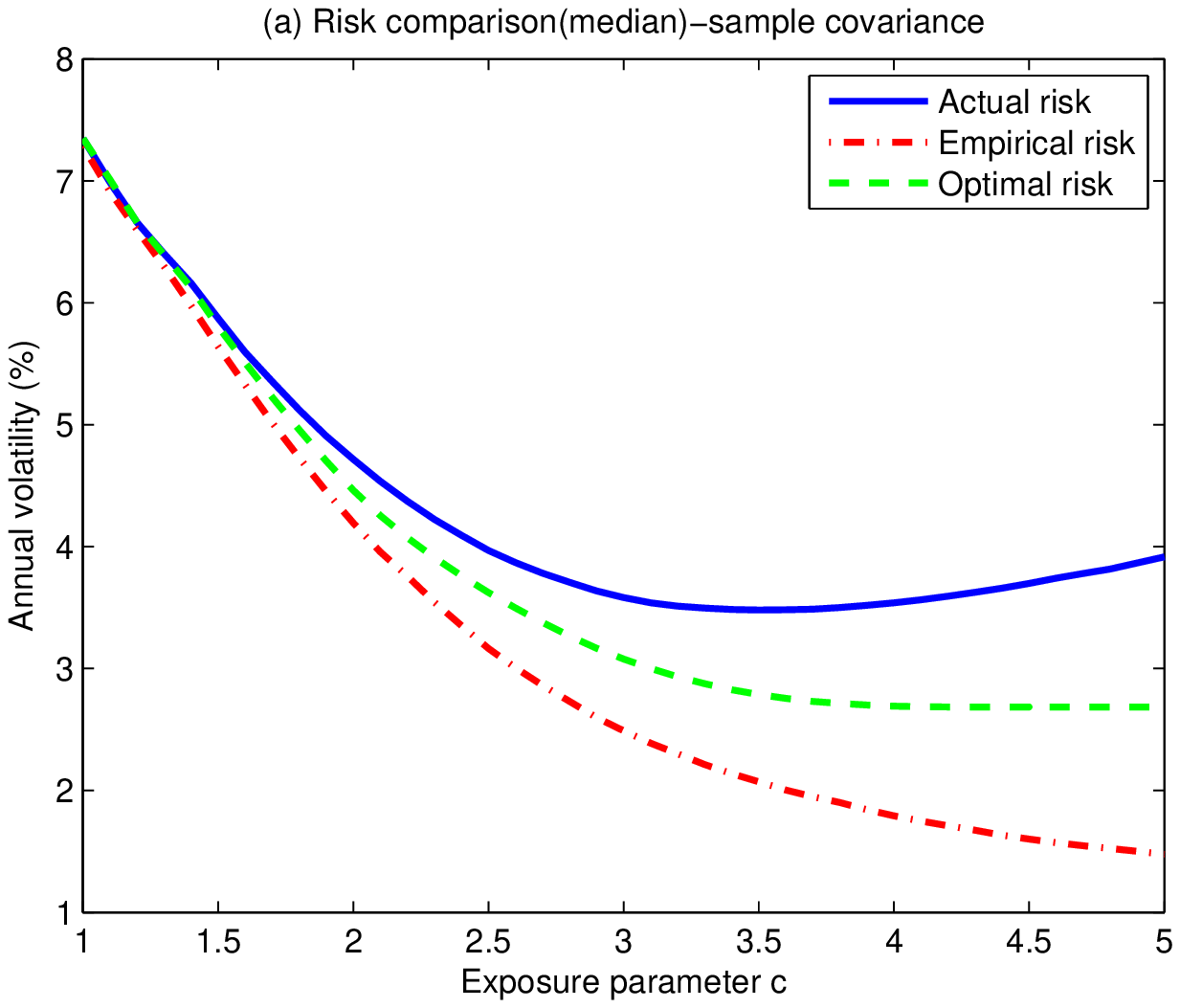} &
   \includegraphics[scale=0.5]{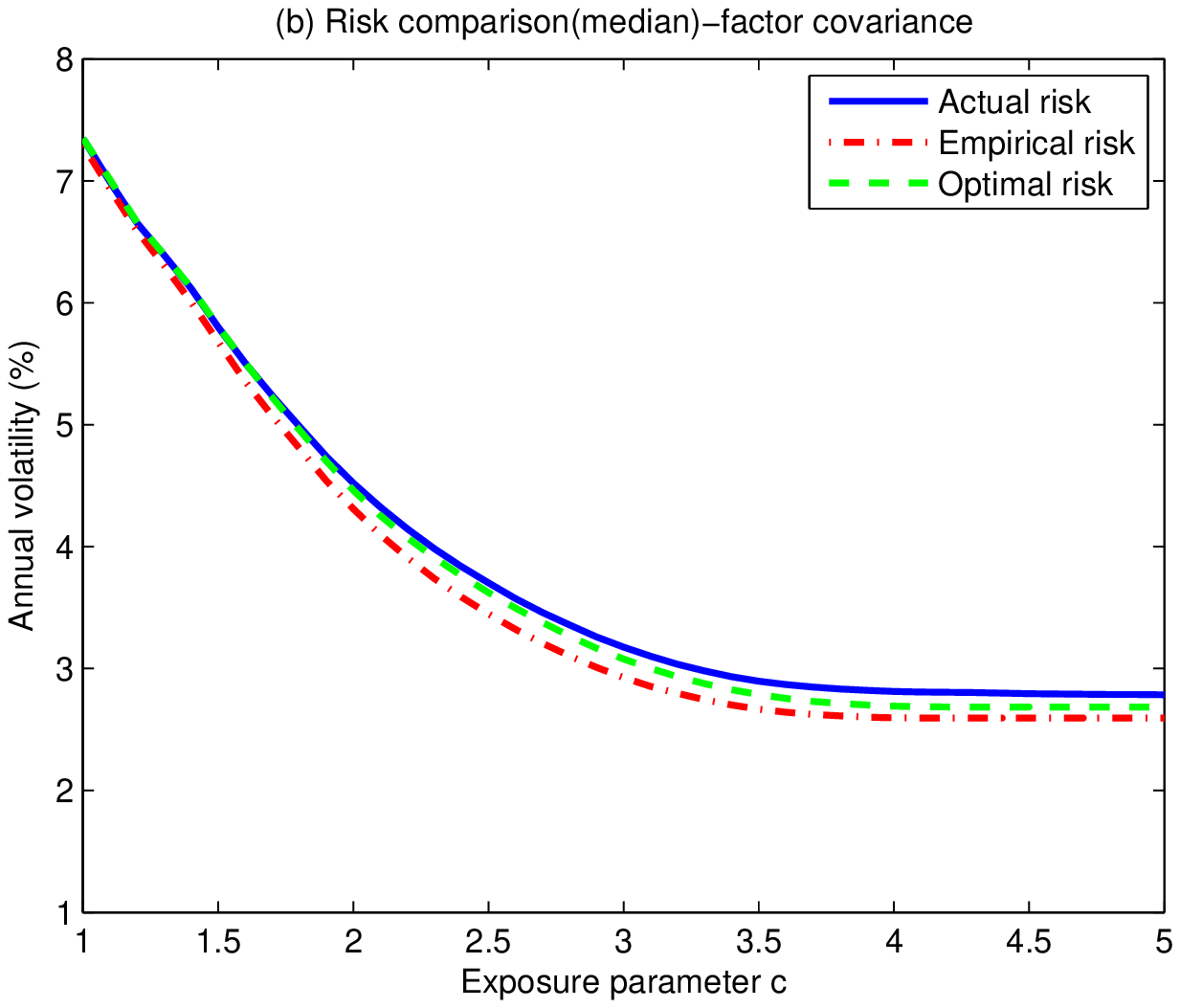}
\end{tabular}
\begin{singlespace}
   \caption{\small The 10\%, 50\% and 90\% quantiles of the actual risks of
   the 101 empirically chosen portfolios for each given gross exposure parameter
   $c$ are shown in (a) (sample covariance matrix) and (b) (factor model)
   for the case with 200 stocks and the daily returns in a year.
   They indicate the sampling variability among 101 simulations.
   The theoretical optimal risk, the median of the
   actual risks and the median of the empirical risks of 101
   empirically selected portfolios are also summarized in (c) (based
   on the sample covariance) and (d) (based on the factor model).
   }
\end{singlespace}
\end{center}
\end{figure}

We first examine the case $p = 200$ with a sample of size 252.
Figure 5(a) summarizes the 10th, 50th and 90th percentiles of the
actual risks of empirically selected portfolios among 101
simulations. They summarize the distributions of the actual risk of
the optimally selected portfolios based on  101 empirically
simulated data sets. The variations are actually small. Table 3
(bottom panel) also includes the theoretical optimal risk, the
median of the actual risks of 101 empirically selected optimal
portfolios, and the median of the empirical risks of those 101
selected portfolios. This part indicates the typical error of the
risk approximations.  It is clear from Figure 5(c) that the
theoretical risk fails to decrease noticeably when $c=3$ and
increasing the gross-exposure constraint will not improve very much
the theoretical optimal risk profile.  On the other hand, increasing
gross exposure $c$ makes it harder to estimate theoretical
allocation vector. As a result, the actual risk increases when $c$
gets larger.

Combining the results in both top and bottom panels, Table 3 gives a
comprehensive overview of the risk approximations.   For example,
when the global exposure parameter is large, the approximation
errors dominate the sampling variability.  It is clear that the risk
approximations are much more accurate for the covariance matrix
estimation based on the factor model. This is somewhat expected as
the data generating process is a factor model: there are no modeling
biases in estimating the covariance matrix. For the sample
covariance estimation, the accuracy is fairly reasonable until the
gross exposure parameter exceeds 2.

\begin{singlespace}
\begin{table}[t]   
\begin{center}
\caption{\bf Empirical and actual risks for selected portfolios}
\end{center}
\vspace*{-0.3 in}

\noindent \small This table is based on 101 simulations. Each
simulation generates 252 daily returns of 200 stocks from the
Fama-French three-factor model. The covariance is estimated by
sample covariance matrix or the factor model (\ref{d3}). The
penalized least-squares (\ref{c3}) is used to construct the optimal
portfolios.

\begin{center}
\begin{tabular}
[c]{clccccccccc} \hline \multicolumn{10}{c}{\em Sample covariance matrix}\\
\hline
  && Theorectical Cov &\hspace{0.2 in} &\multicolumn{6}{c}{Sample covariance} \\
  \cline{3-3} \cline{5-10}
c& \hspace*{0.1 in} &
  Theoretical opt. & & min & $1^{st}$ quantile &
              & median & $3^{rd}$ quantile & max \\ \hline
1   & Actual    & 7.35 & & 7.35  & 7.36   &  & 7.37   & 7.38 & 7.43 \\
    & Empirical & 7.35 & & 6.64  & 7.07   &  & 7.28   & 7.52 & 8.09 \\
2   & Actual    & 4.46 & & 4.48  & 4.64   &  & 4.72   & 4.78 & 5.07  \\
    & Empirical & 4.46 & & 3.71  & 4.04   &  & 4.19   & 4.36 & 4.64   \\
3   & Actual    & 3.07 & & 3.41  & 3.53   &  & 3.58   & 3.66 & 3.84   \\
    & Empirical & 3.07 & & 2.07  & 2.40   &  & 2.49   & 2.60 & 2.84  \\
4   & Actual    & 2.69 & & 3.31  & 3.47   &  & 3.54   & 3.61 & 3.85   \\
    & Empirical & 2.69 & & 1.48  & 1.71   &  & 1.79   & 1.87 & 2.05  \\
5   & Actual    & 2.68 & & 3.62  & 3.81   &  & 3.92   & 3.99 & 4.25 \\
    & Empirical & 2.68 & & 1.15  & 1.41   &  & 1.48   & 1.57 & 1.73  \\
\hline
\multicolumn{10}{c}{\em Factor-based covariance matrix}\\
\hline
1   & Actual       & 7.35 & &   7.35    & 7.36    & & 7.37    & 7.39   & 7.41 \\
    & Empirical    & 7.35 & &   6.60    & 7.07    & & 7.29    & 7.50   & 8.07 \\
2   & Actual       & 4.46 & &   4.46    & 4.48    & & 4.52    & 4.57   & 4.74 \\
    & Empirical    & 4.46 & &   3.96    & 4.19    & & 4.31    & 4.45   & 4.80 \\
3   & Actual       & 3.07 & &   3.14    & 3.16    & & 3.18    & 3.19   & 3.26 \\
    & Empirical    & 3.07 & &   2.75    & 2.86    & & 2.93    & 2.98   & 3.18 \\
4   & Actual       & 2.69 & &   2.76    & 2.79    & & 2.81    & 2.83   & 2.90 \\
    & Empirical    & 2.69 & &   2.49    & 2.56    & & 2.60    & 2.63   & 2.75 \\
5   & Actual       & 2.68 & &   2.73    & 2.77    & & 2.78    & 2.80   & 2.87 \\
    & Empirical    & 2.68 & &   2.49    & 2.56    & & 2.59    & 2.62   & 2.74 \\
\hline
\end{tabular}
\end{center}
\end{table}
\end{singlespace}

Table 3 furnishes some additional details for Figure 5.  For the
optimal portfolios with no-short-sale constraint, the theoretical
and empirical risks are very close to each other.  For the global
minimum variance portfolio, which corresponds to a large $c$, the
empirical and actual risks of an empirically selected portfolio can
be quite different. The allocation vectors based on a known
covariance matrix can also be very different from that based on the
sample covariance.  To help gauge the scale, we note that for the
true covariance, the global minimum variance portfolio has $c =
4.22$, which involves 161\% of short positions, and minimum risk
2.68\%.

\begin{figure}[t]    
\begin{center}
\begin{tabular}{c c}
   \includegraphics[scale=0.5]{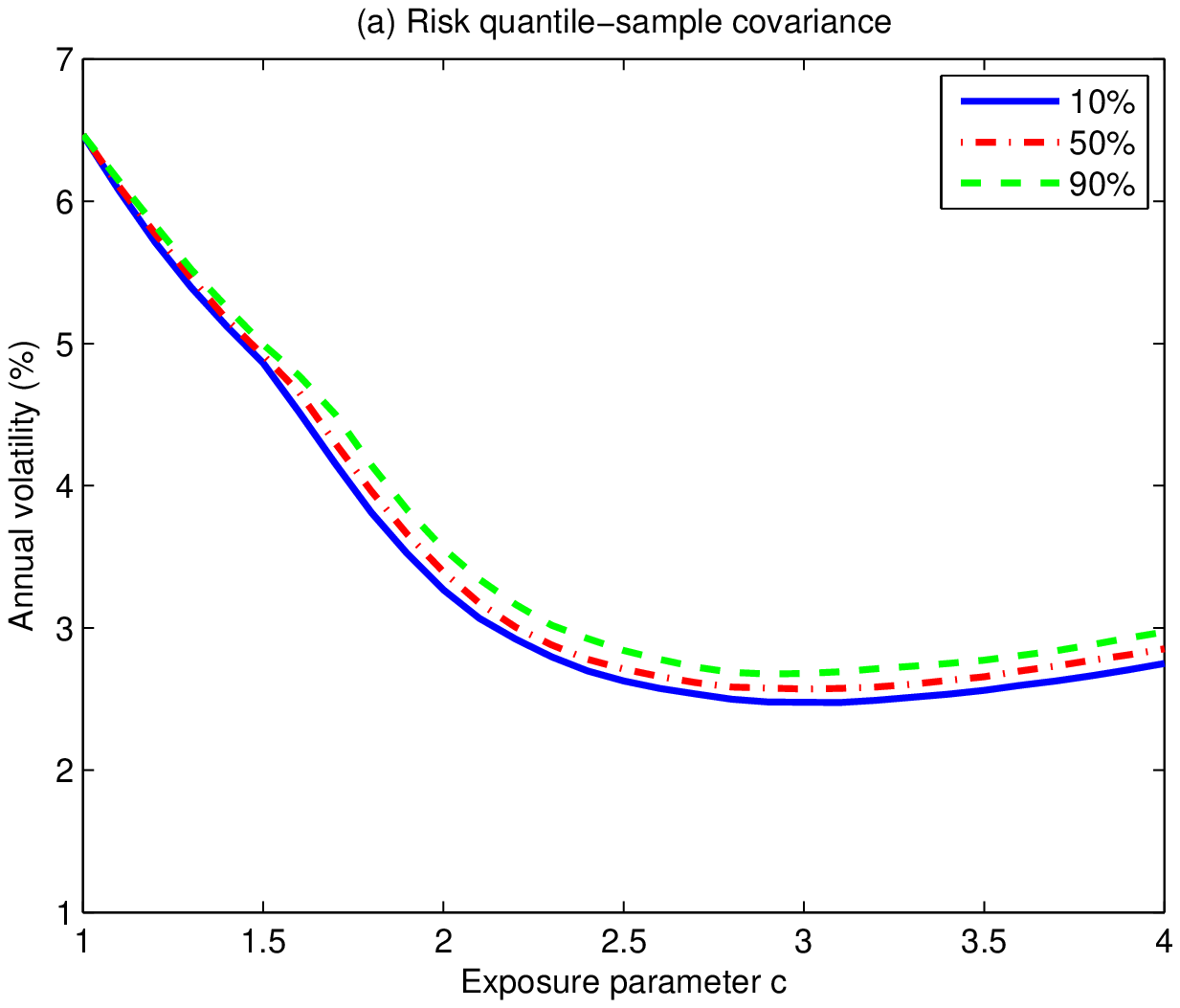} &
   \includegraphics[scale=0.5]{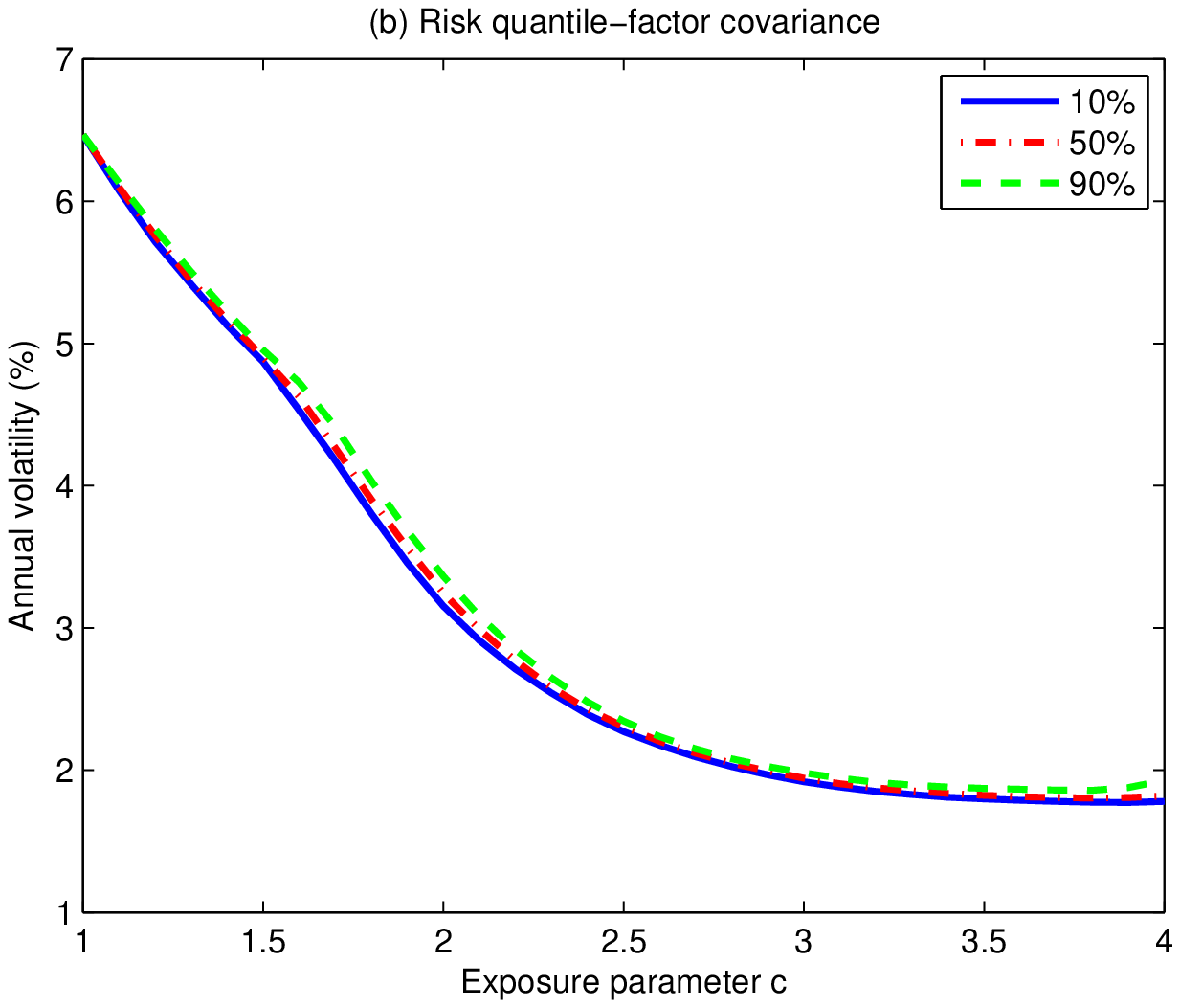} \\
   \includegraphics[scale=0.5]{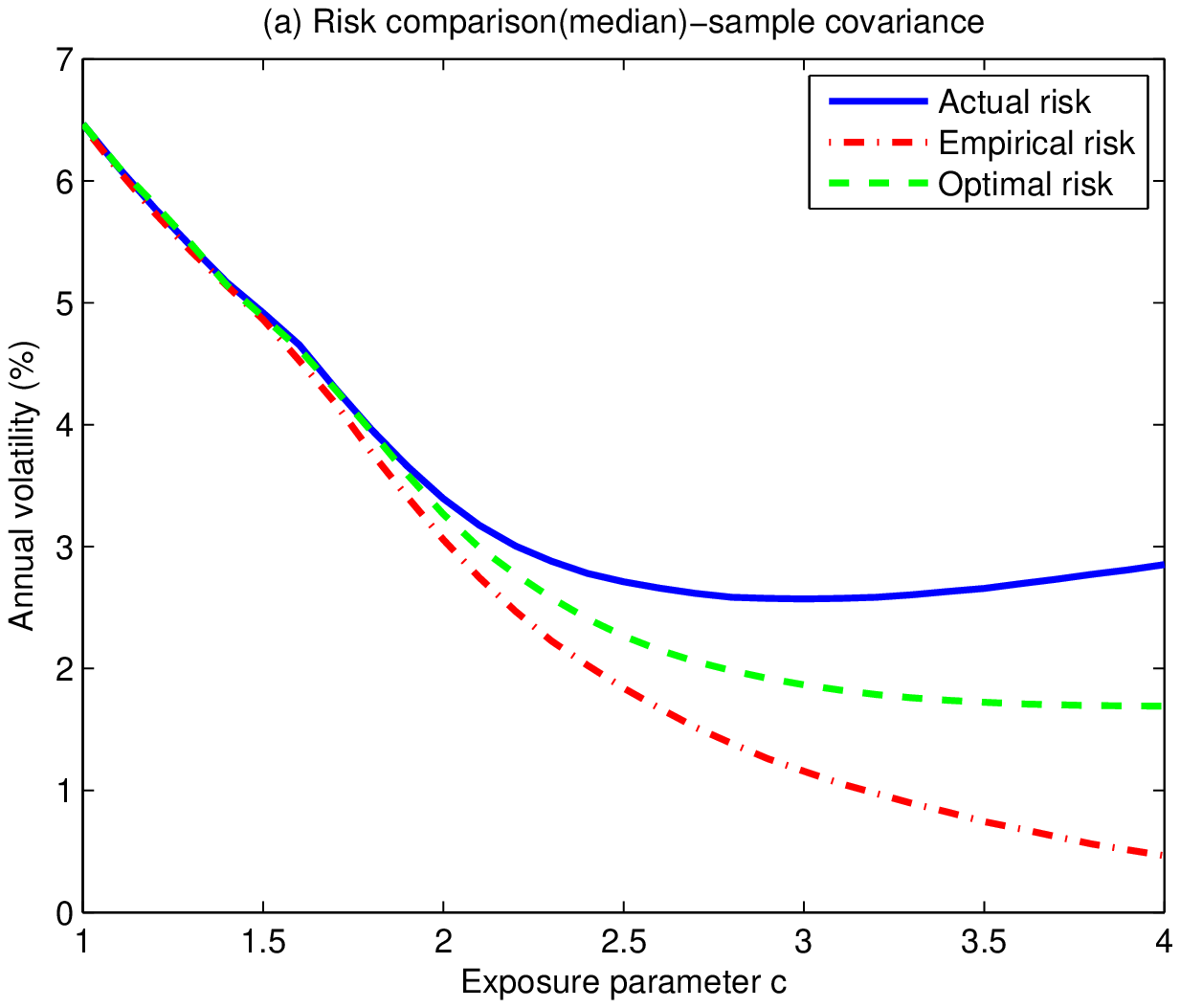} &
  \includegraphics[scale=0.5]{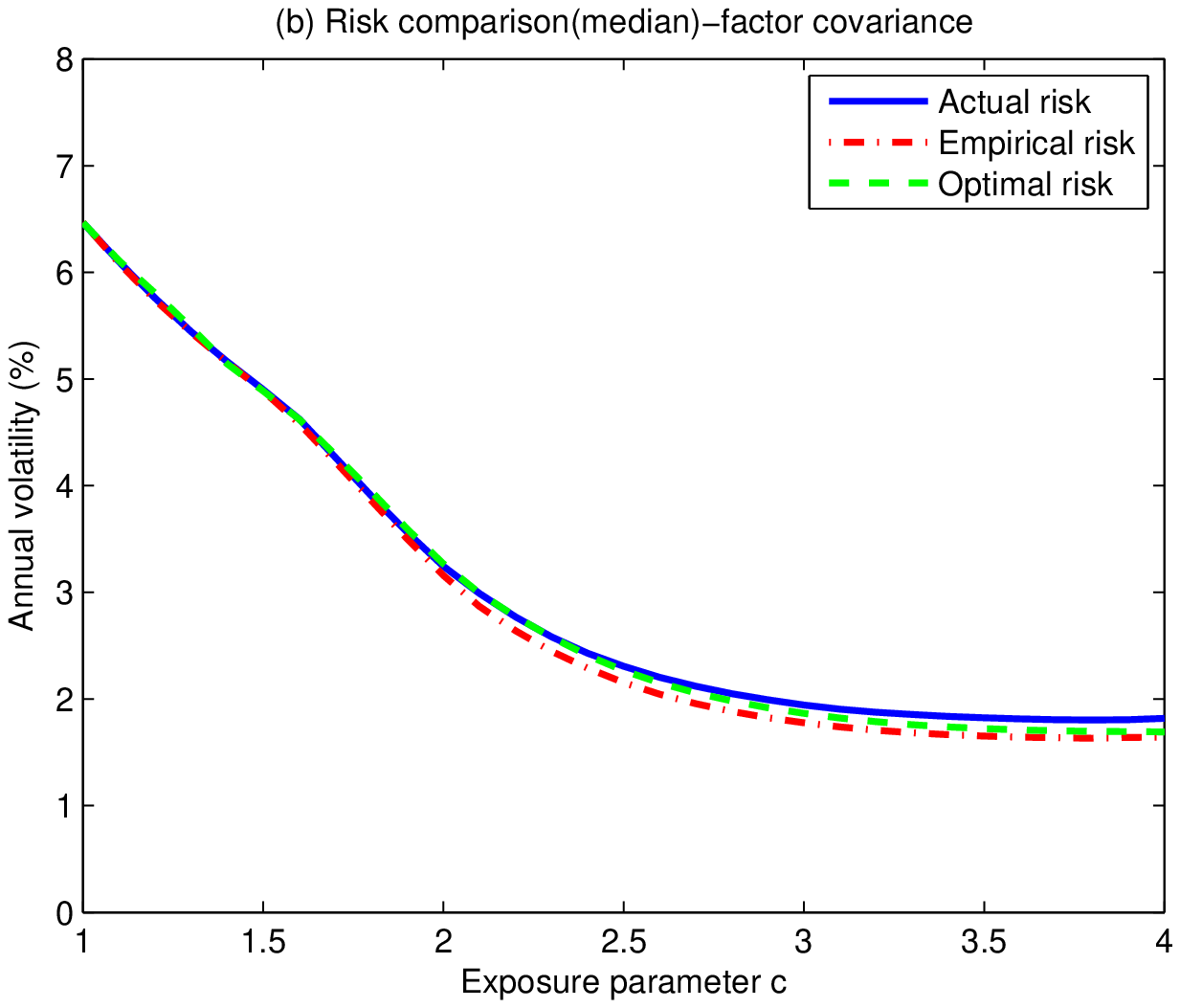}
\end{tabular}
\begin{singlespace}
   \caption{\small This is similar to Figure 5 except $p=500$.  The
   sample covariance matrix is always degenerate under this setting
   ($n=252$).  Nevertheless, for the given range of $c$, the
   gross-constrained portfolio performs normally.
   The same captions as Figure 5 are used.
   }
\end{singlespace}
\end{center}
\end{figure}

\begin{singlespace}
\begin{table}[t]             
\begin{center}
\caption{\bf Empirical and actual risks for selected portfolios}
\end{center}
\vspace*{-0.3 in}

\noindent \small This is a similar to Table 3 except $p=500$. In
this case, the sample covariance matrix is always degenerate.

\begin{center}
\begin{tabular}
[c]{clccccccccc} \hline \multicolumn{10}{c}{\em Sample covariance matrix}\\
\hline
  && Theoretical Cov &\hspace{0.2 in} &\multicolumn{6}{c}{Sample covariance} \\
  \cline{3-3} \cline{5-10}
c& \hspace*{0.1 in} &
  Theoretical opt. & & min & $1^{st}$ quantile &
              & median & $3^{rd}$ quantile & max \\ \hline

1   & Actual    & 6.47 & & 6.47 & 6.48 & & 6.49 & 6.50 & 6.53     \\
    & Empirical & 6.47 & & 5.80 & 6.28 & & 6.45 & 6.67 & 7.13     \\
2   & Actual    & 3.27 & & 3.21 & 3.29 & & 3.39 & 3.47 & 3.73     \\
    & Empirical & 3.27 & & 2.54 & 2.92 & & 3.06 & 3.22 & 3.42     \\
3   & Actual    & 1.87 & & 2.42 & 2.53 & & 2.57 & 2.63 & 2.81     \\
    & Empirical & 1.87 & & 0.88 & 1.09 & & 1.15 & 1.24 & 1.49     \\
4   & Actual    & 1.69 & & 2.65 & 2.79 & & 2.85 & 2.92 & 3.21     \\
    & Empirical & 1.69 & & 0.24 & 0.41 & & 0.46 & 0.52 & 0.77     \\

\hline
\multicolumn{10}{c}{\em Factor-based covariance matrix}\\
\hline

1   & Actual      & 6.47 & & 6.47 & 6.48 & & 6.49 & 6.51 & 6.55 \\
    & Empirical   & 6.47 & & 5.80 & 6.29 & & 6.45 & 6.67 & 7.15 \\
2   & Actual      & 3.27 & & 3.16 & 3.21 & & 3.35 & 3.39 & 3.48  \\
    & Empirical   & 3.27 & & 2.74 & 3.02 & & 3.16 & 3.29 & 3.52   \\
3   & Actual      & 1.87 & & 1.91 & 1.93 & & 1.94 & 1.96 & 2.02   \\
    & Empirical   & 1.87 & & 1.70 & 1.75 & & 1.78 & 1.81 & 1.89  \\
4   & Actual      & 1.69 & & 1.75 & 1.79 & & 1.82 & 1.85 & 2.87   \\
    & Empirical   & 1.69 & & 1.59 & 1.63 & & 1.64 & 1.67 & 2.75  \\
\hline
\end{tabular}
\end{center}
\end{table}
\end{singlespace}

We now consider the case where there are 500 potential stocks with
only a year of data ($n=252$).  In this case, the sample covariance
matrix is always degenerate.  Therefore, the global minimum
portfolio based on empirical data is meaningless, which always has
empirical risk zero.  In other words, the difference between the
actual and empirical risks of such an empirically constructed global
minimum portfolio is substantial.  On the other hand, with the
gross-exposure constraint, the actual and empirical risks
approximate quite well for a wide range of gross exposure
parameters. To gauge the relative scale of the range, we note that
for the given theoretical covariance, the global minimum portfolio
has $c =4.01$, which involves 150\% of short positions with the
minimal risk 1.69\%.

The sampling variability for the case with 500 stocks is smaller
than the case that involves 200 stocks, as demonstrated in Figures 5
and 6. The approximation errors are also smaller.  These are due to
the fact that with more stocks, the selected portfolio is generally
more diversified and hence the risks are generally smaller.  The
optimal no-short-sale portfolio, selected from 500 stocks, has
actual risk 6.47\%, which is not much smaller than 7.35\% selected
from 200 stocks.  As expected, the factor-based model has a better
estimation accuracy than that based on the sample covariance.

\section{Empirical Studies}
\subsection{Fama-French 100 Portfolios}

\begin{singlespace}
\begin{table}[h]             
\begin{center}
\caption{\bf Returns and Risks based on 100 Fama-French Industrial}
\end{center}
\vspace*{-0.3 in}

\noindent\small We use the daily returns of 100 industrial
portfolios formed by size and book to market from the website of
Kenneth French from Jan 2, 1998 to December 31, 2007.  At the end of
each month from 1998 to 2007, the covariance of the 100 assets is
estimated according to various estimators using the past 12 months'
daily return data. We use these covariance matrices to construct
optimal portfolios with various exposure constraints. We hold the
portfolios for one month. The means, standard deviations and other
characteristics of these portfolios are recorded. (NS: no short
sales portfolio; GMV: Global minimum variance portfolio)
\begin{center}
\small \begin{tabular}{lccccccc} \hline
& Mean & Std Dev  & Sharpe & Max    & Min    & No. of Long &  No. of Short \\
Methods & \% & \% &  Ratio & Weight & Weight &  Positions &Positions
 \\ \hline
\multicolumn{8}{c}{\em {Sample Covariance Matrix Estimator}} \\
\hline
 No short(c = 1)       &19.51  & 10.14 & 1.60  & 0.27 & -0.00 & 6  &0  \\
 Exact(c = 1.5)        &21.04  & 8.41  & 2.11  & 0.25 & -0.07 & 9  &6  \\
 Exact(c = 2)          &20.55  & 7.56  & 2.28  & 0.24 & -0.09 & 15 &12  \\
 Exact(c = 3)          &18.26  & 7.13  & 2.09  & 0.24 & -0.11 & 27 &25  \\
 Approx. (c = 2, Y=NS) &21.16  & 7.89  & 2.26  & 0.32 & -0.08 & 9  &13   \\
 Approx. (c = 3, Y=NS) &19.28  & 7.08  & 2.25  & 0.28 & -0.11 & 23 &24   \\
 GMV Portfolio         &17.55  & 7.82  & 1.82  & 0.66 & -0.32 & 52 &48  \\
 \hline

 \multicolumn{8}{c}{\em {Factor-Based  Covariance Matrix Estimator}} \\
 \hline
 No short(c = 1)      &20.40  & 10.19 & 1.67 &  0.21 & -0.00  &7 &0  \\
 Exact(c = 1.5)       &22.05  & 8.56 & 2.19 &  0.19 & -0.05  &11 &8  \\
 Exact(c = 2)         &21.11  & 7.96 & 2.23 &  0.18 & -0.05  &17 &18 \\
 Exact(c = 3)         &19.95  & 7.77 & 2.14 &  0.17 & -0.05  &35 &41 \\
 Approx. (c=2, Y=NS)  &21.71  & 8.07 & 2.28 &  0.24 & -0.04  &10 &19 \\
 Approx. (c=3, Y=NS) &20.12  & 7.84 & 2.14 &  0.18 & -0.05  &33 &43  \\
 GMV Portfolio        &19.90  & 7.93 & 2.09 &  0.43 & -0.14  &45 &55 \\
 \hline

\multicolumn{8}{c}{\em {Covariance Estimation from Risk Metrics}}\\
 \hline
 No short(c = 1)      &15.45 & 9.27 & 1.31 &  0.30 & -0.00 &6  & 0  \\
 Exact(c = 1.5)       &15.96 & 7.81 & 1.61 &  0.29 & -0.07 &9  & 5  \\
 Exact(c = 2)       &14.99 & 7.38 & 1.58 &  0.29 & -0.10 &13 & 9  \\
 Exact(c = 3)         &14.03 & 7.34 & 1.46 &  0.29 & -0.13 &21 & 18 \\
 Approx. (c=2, Y=NS)  &15.56 & 7.33 & 1.67 &  0.34 & -0.08 &9  & 11 \\
 Approx. (c=3, Y=NS)  &15.73 & 6.95 & 1.78 &  0.30 & -0.11 &20 & 20 \\
 GMV Portfolio        &13.99 & 9.47 & 1.12 &  0.78 & -0.54 &53 & 47 \\
\hline

\multicolumn{8}{c}{\em {Unmanaged Index}} \\ \hline
 Equal weighted           &  10.86   & 16.33    & 0.46   & 0.01   & 0.01   & 100   & 0 \\
 CRSP                     &  8.2     & 17.9     & 0.26   &&&& \\
\hline
\end{tabular}
\end{center}
\end{table}
\end{singlespace}

\begin{figure}[t]   
\begin{center}
\begin{tabular}{c c}
    \includegraphics[scale=0.5]{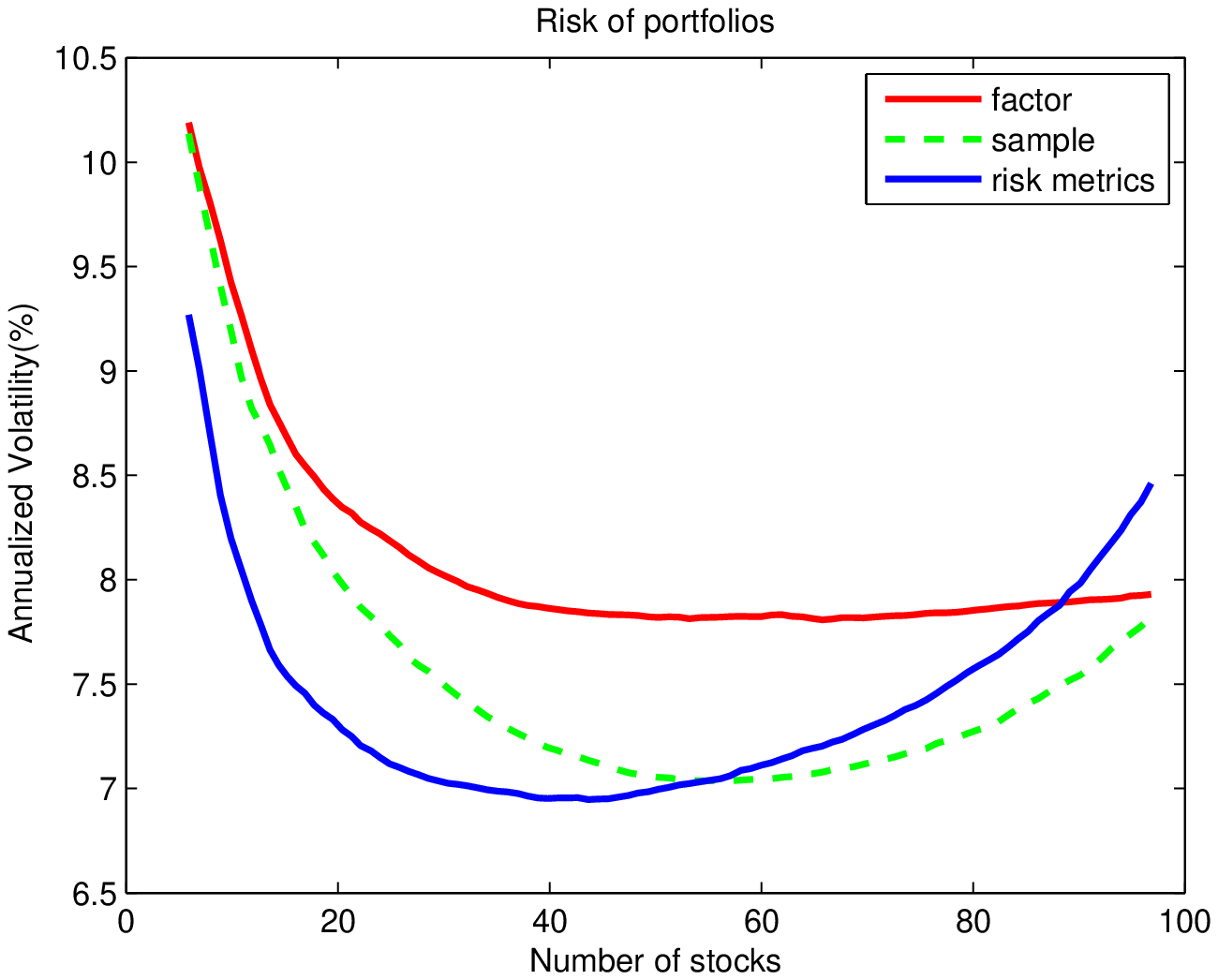} &
    \includegraphics[scale=0.5]{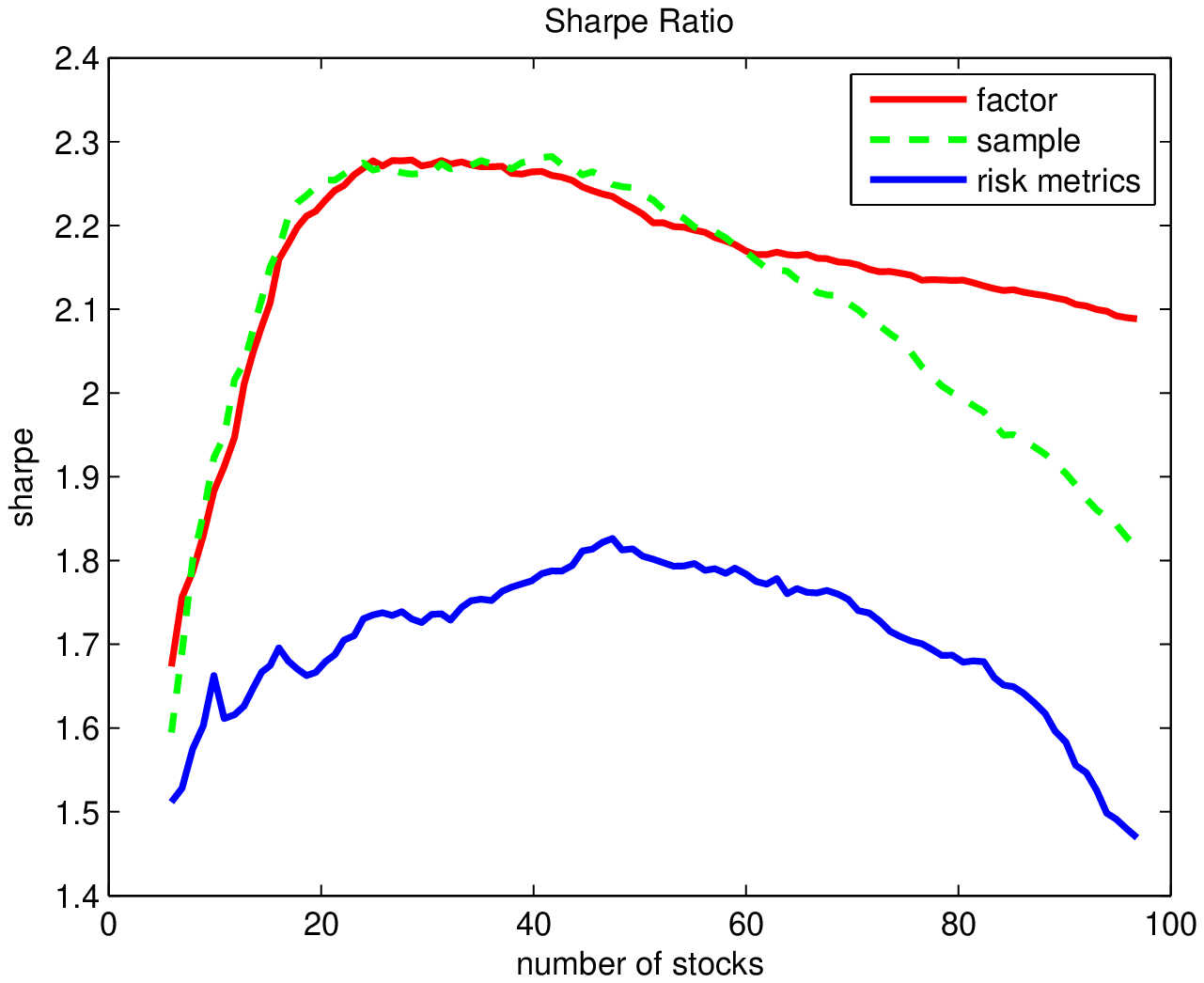} \\
    \includegraphics[scale=0.5]{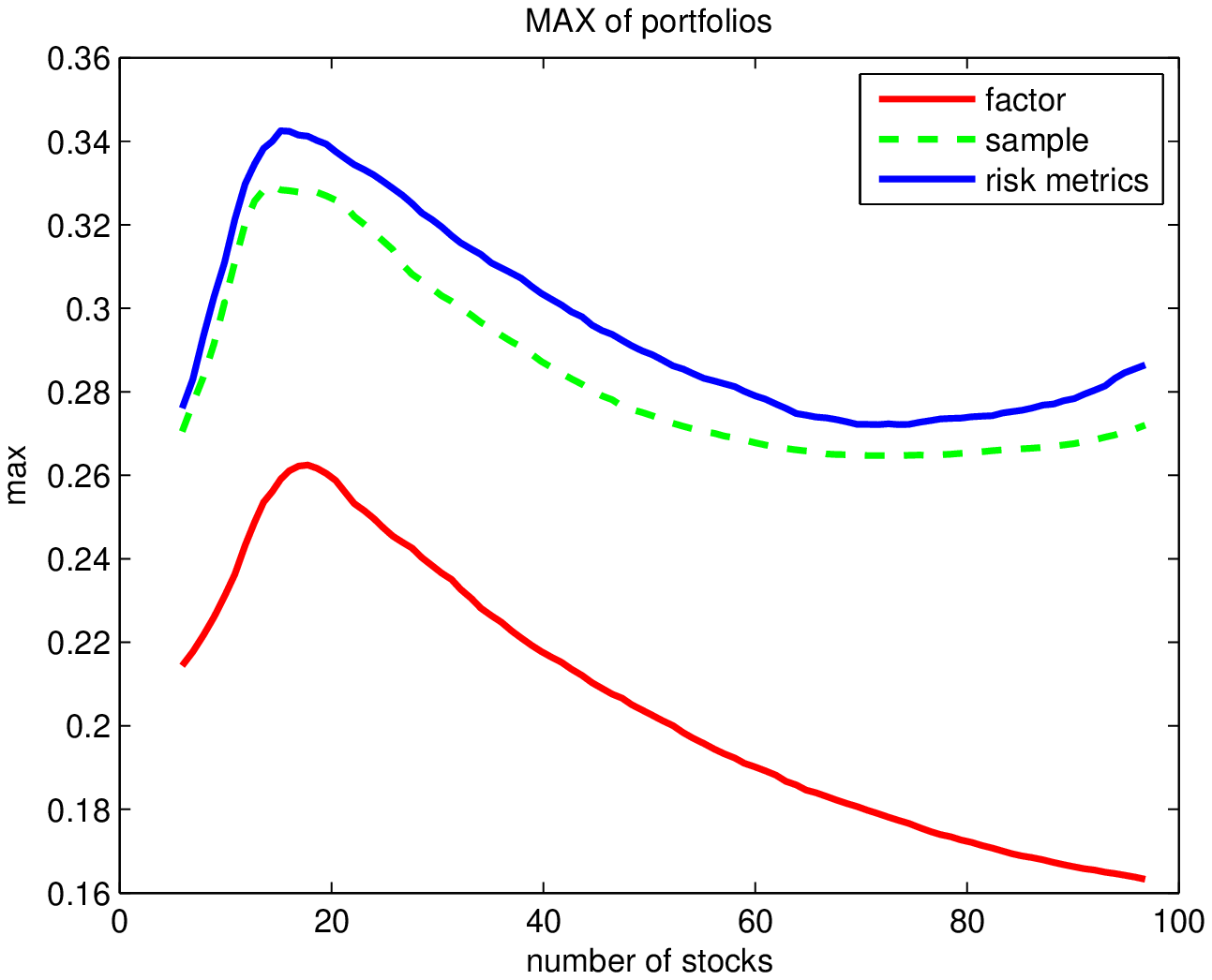} &
    \includegraphics[scale=0.5]{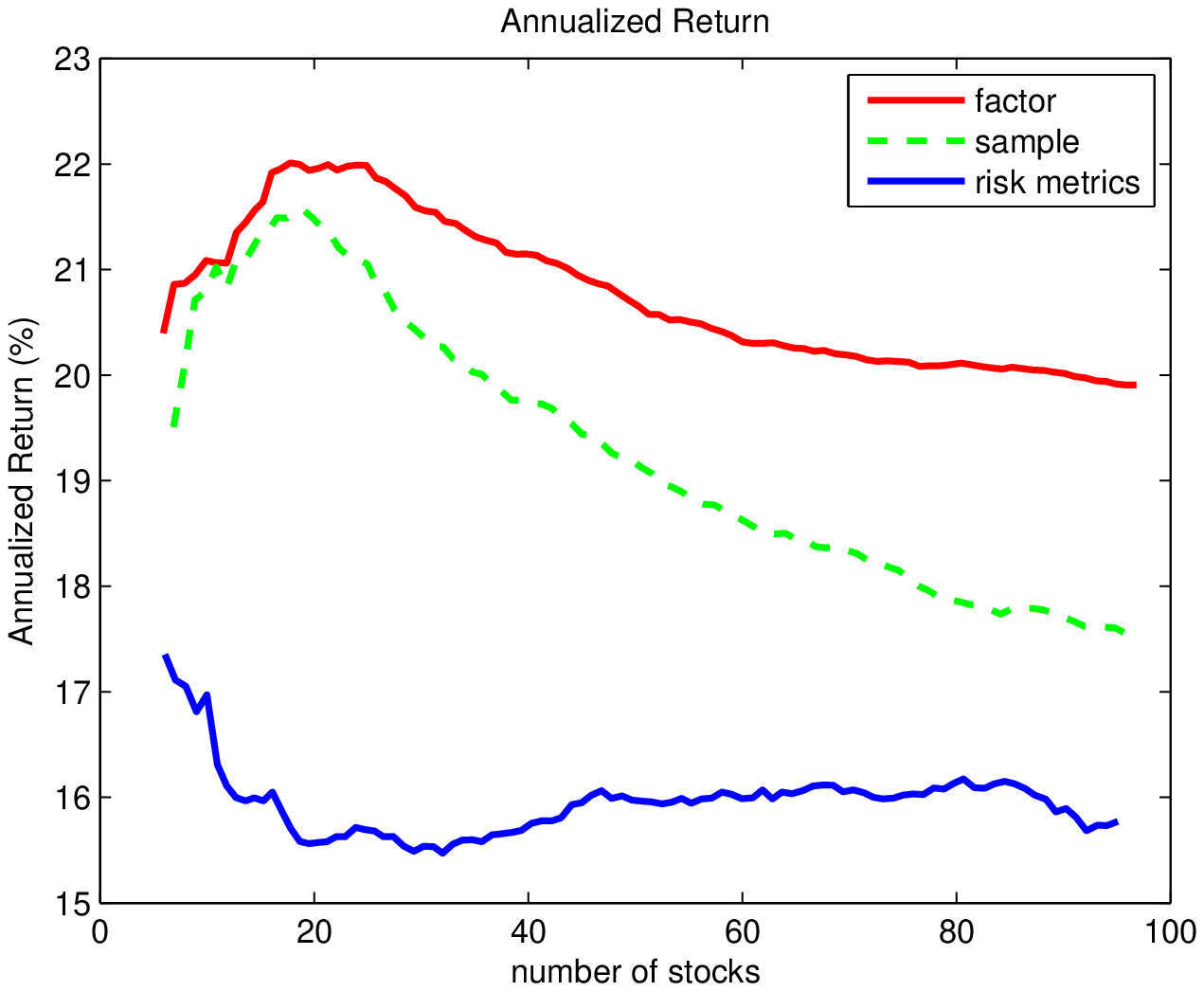} \\
\end{tabular}
\begin{singlespace}
   \caption{Characteristics of invested portfolios as a function of number of
   assets from the Fama-French 100 industrial portfolios formed by
   the size and book to market from Jan 2, 1998 to December 31,
   2007.
    (a) Annualized risk of portfolios. (b)Sharpe ratio of portfolios.
   (c)Max weight of allocations. (d)Annualized return of portfolios}
\end{singlespace}
\end{center}
\end{figure}

We use the daily returns of 100 industrial portfolios formed on size
and book to market ratio from the website of Kenneth French from Jan
2, 1998 to December 31, 2007.  These 100 portfolios are formed by
the two-way sort of the stocks in the CRSP database, according to
the market equity and the ratio of book equity to market equity, 10
categories in each variable.  At the end of each month from 1998 to
2007, the covariance matrix of the 100 assets is estimated according
to three estimators, the sample covariance, Fama-French 3-factor
model, and the RiskMetrics with $\lambda = 0.97$, using the past 12
months' daily return data. These covariance matrices are then used
to construct optimal portfolios under various exposure constraints.
The portfolios are then held for one month and rebalanced at the
beginning of the next month. The means, standard deviations and
other characteristics of these portfolios are recorded and presented
in Table 5. They represent the actual returns and actual risks.
Figure 7, produced by using the LARS-LASSO algorithm, provides some
additional details to these characteristics in terms of the number
of assets held.  The optimal portfolios with the gross-exposure
constraints pick certain numbers of assets each month.  The average
numbers of assets over the study period are plotted in the x-axis.

First of all, the optimal no-short-sale portfolios, while selecting
about 6 assets from 100 portfolios, are not diversified enough.
Their risks can easily be improved by relaxing the gross-exposure
constraint with $c=2$ that has 50\% short positions and 150\% long
positions. This is shown in Table 5 and Figure 7(a), no matter which
method is used to estimate the covariance matrix.  The risk stops
decreasing dramatically once the number of stocks exceeds 20.
Interestingly, the Sharpe ratios peak around 20 stocks too. After
that point, the Sharpe Ratio actually falls for the covariance
estimation based on the sample covariance and the factor model.

The portfolios selected by using the RiskMetrics have lower risks.
In comparison with the sample covariance matrix, the RiskMetrics
estimates the covariance matrix using a much smaller effective time
window. As a result, the biases are usually smaller than the sample
covariance matrix.  Since each asset is a portfolio in this study,
its risk is smaller than stocks.   Hence, the covariance matrix can
be estimated more accurately with the RiskMetrics in this study.
This explains why the resulting selected portfolios by using
RiskMetrics have smaller risks.   However, their associated returns
tend to be smaller too. As a result, their Sharpe ratios are
actually smaller. The Sharpe ratios actually peak at around 50
assets.

It is surprising to see that the unmanaged equally weighted
portfolio, which invests 1 percent on each of the 100 industrial
portfolios, is far from optimal in terms of the risk during the
study period.   The value-weighted index CRSP does not fare much
better.   They are all outperformed by the optimal portfolios with
gross-exposure constraints during the study period.  This is in line
with our theory.  Indeed, the equally weighted portfolio and CRSP
index are two specific members of the no-short-sale portfolio, and
should be outperformed by the optimal no-short-sale portfolio, if
the covariance matrix is estimated with reasonable accuracy.

From Table 5, it can also be seen that our approximation algorithm
yields very close solution to the exact algorithm. For example,
using the sample covariance matrix,  the portfolios constructed
using the exact algorithm with $c=3$ has the standard deviation of
7.13\%, whereas the portfolios constructed using the approximate
algorithm has the standard deviation of 7.08\%.   In terms of the
average numbers of selected stocks over the 10-year study period,
they are close too.

\subsection{Russell 3000 Stocks}

\begin{singlespace}
\begin{table}[h]             
\begin{center}
\caption{\bf Returns and Risks based on random 400 portfolio}
\end{center}
\vspace*{-0.3 in}

\noindent\small We pick 1000 stocks from Russell 3000 with least
percents of missing data from Jan 2, 2003 to December 31, 2007.
Among the 1000 stocks, we randomly pick 400 stocks to avoid survival
bias. At the end of each month from 2003 to 2007, the covariance of
the 400 stocks is estimated according to various estimators using
the past 24 months' daily return data. We use these covariance
matrices to construct optimal portfolios under various
gross-exposure constraints. We hold the portfolio for one month. The
standard deviations and other characteristics of these portfolios
are recorded. (NS: no short sales; MKT: return of S\&P 500 index;
GMV: Global minimum variance portfolio)
\begin{center}
\small \begin{tabular}{lccccccc} \hline
 & Std Dev & Max    & Min    & No. of Long &  No. of Short \\
Methods  & \%  & Weight & Weight &  Positions &Positions
 \\ \hline
\multicolumn{6}{c}{\em {Sample Covariance Matrix Estimator}} \\
\hline
 No short                  & 9.72   & 0.17 & -0.00 & 51  &0  \\
 Approx (NS,  c= 1.5)      & 8.85   & 0.21 & -0.06 & 54  &33  \\
 Approx (NS,  c= 2)        & 8.65   & 0.19 & -0.07 & 83  &62  \\
 Approx (NS,  c= 2.5)      & 8.62   & 0.17 & -0.08 & 111 &84  \\
 Approx (NS,  c= 3)        & 8.80   & 0.16 & -0.08 & 131 &103  \\
 Approx (NS,  c= 3.5)      & 9.08   & 0.15 & -0.09 & 149 &120  \\
 Approx (MKT, c =1.5)      & 8.79   & 0.15 & -0.08 & 61  &42  \\
 Approx (MKT, c =2)        & 8.64   & 0.15 & -0.08 & 87  &66   \\
 Approx (MKT, c =2.5)      & 8.69   & 0.15 & -0.09 & 109 &88   \\
 Approx (MKT, c =3)        & 8.87   & 0.14 & -0.09 & 128 &108  \\
 Approx (MKT, c =3.5)      & 9.08   & 0.14 & -0.10 & 143 &124  \\
 GMV portfolio             & 14.40  & 0.26 & -0.27 & 209 &191  \\
 \hline

\multicolumn{6}{c}{\em {Factor-Based  Covariance Matrix Estimator}}
\\ \hline
 No short                    & 9.48   & 0.17 & -0.00 & 51  &0\\
 Approx (NS, c= 1.5)         & 8.57   & 0.20 & -0.06 & 54  &36\\
 Approx (NS, c= 2)           & 8.72   & 0.13 & -0.05 & 123 &94\\
 Approx (NS, c= 2.5)         & 9.09   & 0.08 & -0.05 & 188 &159 \\
 Approx (MKT, c =1.5)        & 8.84   & 0.13 & -0.06 & 73  &43\\
 Approx (MKT, c =2)          & 8.87   & 0.10 & -0.05 & 126 &94 \\
 Approx (MKT, c =2.5)        & 9.07   & 0.08 & -0.04 & 189 &164  \\
 GMV portfolio               & 9.23   & 0.08 & -0.05 & 212 &188  \\
 \hline

\multicolumn{6}{c}{\em {Covariance Estimation from Risk Metrics}}\\
\hline
 No short                   & 10.64 &  0.54 & -0.00 & 27 &0\\
 Approx (NS, c= 1.5)        & 10.28 &  0.56 & -0.05 & 38 &25\\
 Approx (NS, c= 2)          & 8.73  &  0.23 & -0.08 & 65 &43\\
 Approx (NS, c= 2.5)        & 8.58  &  0.17 & -0.08 & 94 &67 \\
 Approx (NS, c= 3)          & 8.71  &  0.16 & -0.09 & 119 &90  \\
 Approx (NS, c= 3.5)        & 9.04  &  0.15 & -0.10 & 139 &107  \\
 Approx (MKT, c =1.5)       & 8.70  &  0.27 & -0.15 & 34 &29\\
 Approx (MKT, c =2)         & 8.63  &  0.17 & -0.12 & 60 &49\\
 Approx (MKT, c =2.5)       & 8.58  &  0.14 & -0.12 & 89 &74 \\
 Approx (MKT, c =3)         & 8.65  &  0.15 & -0.12 & 111 &97 \\
 Approx (MKT, c =3.5)       & 8.88  &  0.15 & -0.13 & 131 &114 \\
 GMV portfolio             & 14.67  &  0.27 & -0.27 & 209 &191 \\

\hline
\end{tabular}
\end{center}
\end{table}
\end{singlespace}

\begin{figure}[t]    
\begin{center}
\begin{tabular}{c c}
\includegraphics[scale=0.5]{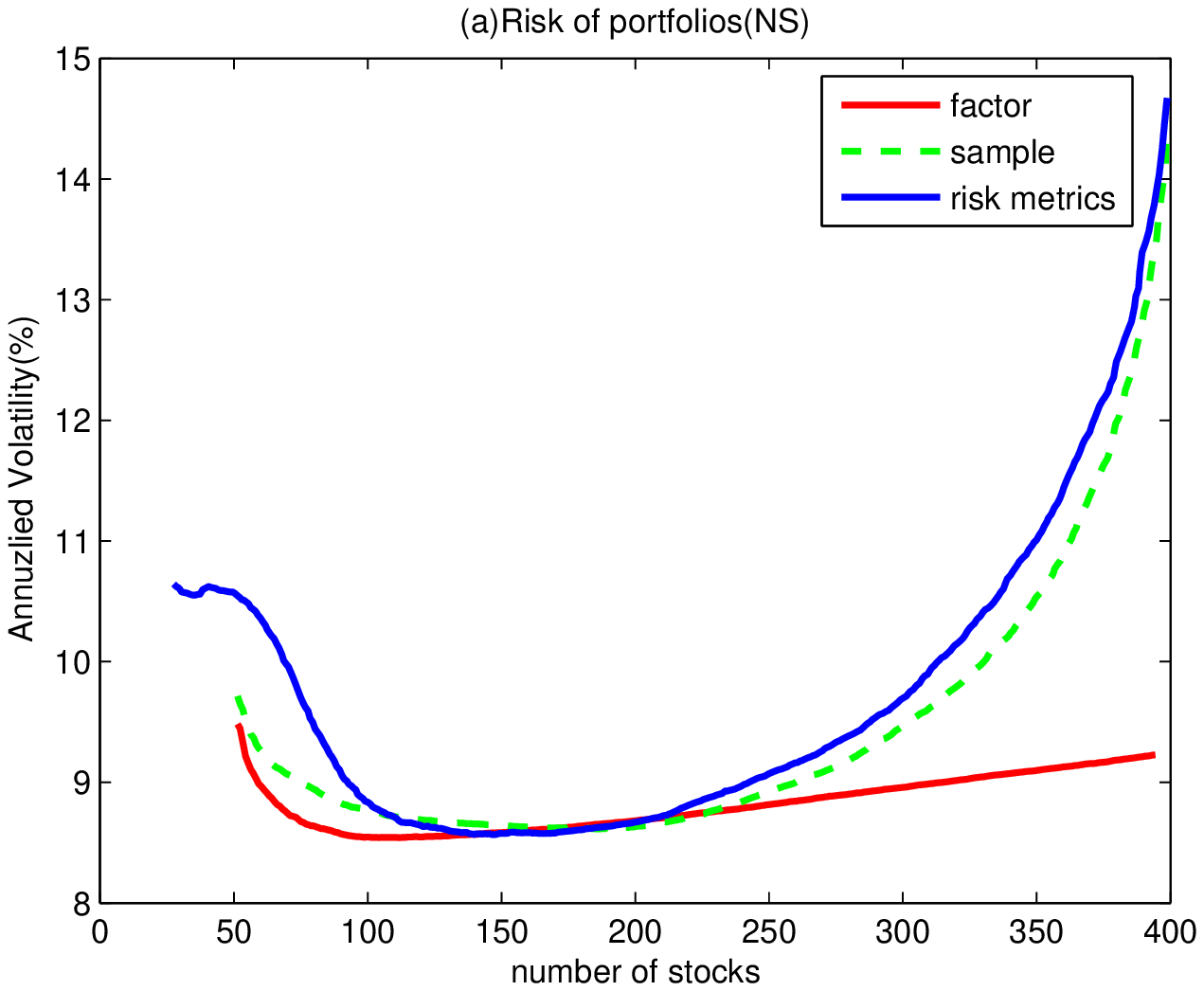} &
   \includegraphics[scale=0.5]{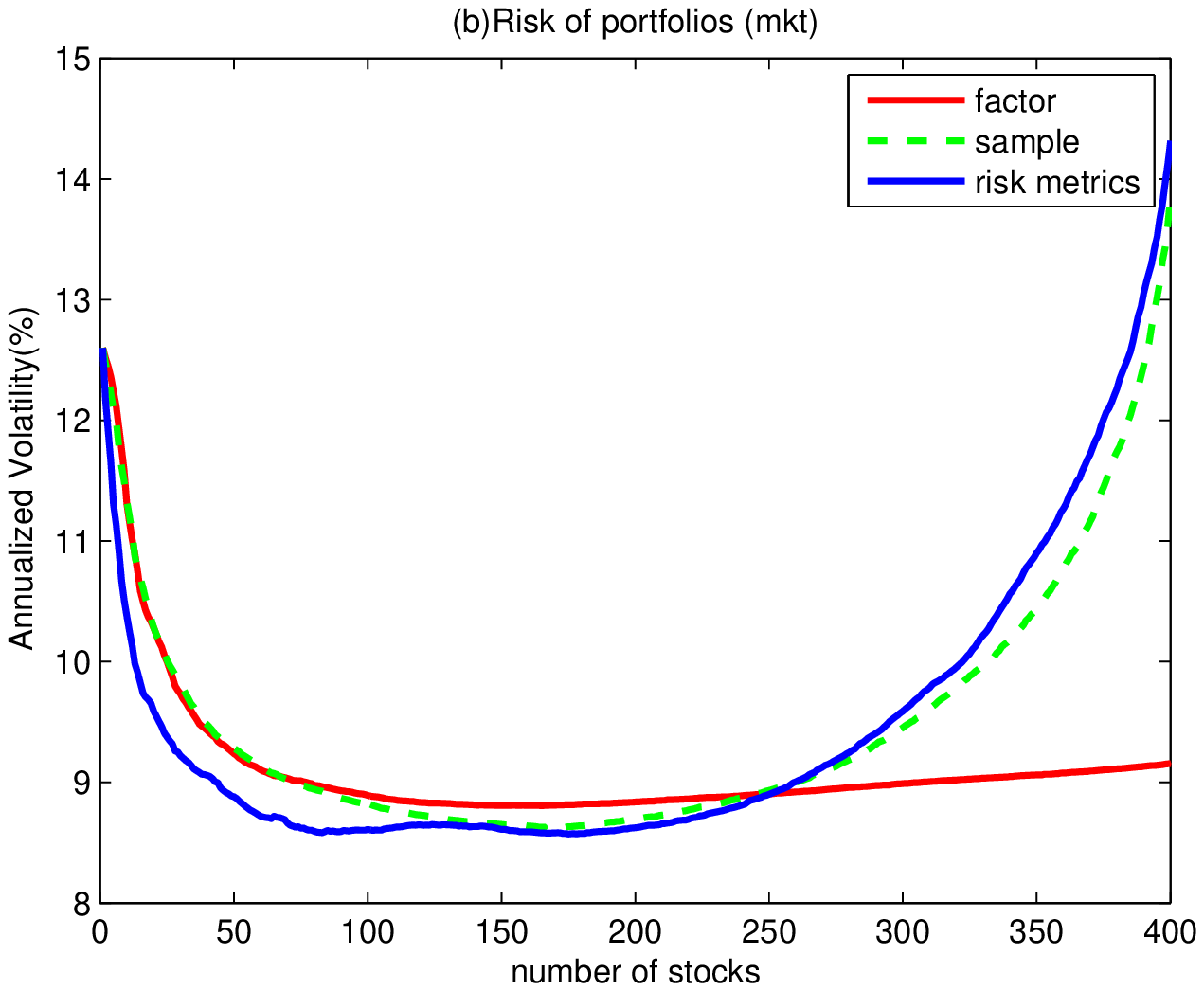} \\
\end{tabular}
\begin{singlespace}
   \caption{Risks of the optimal portfolios as a function of number of stocks for the 400
   randomly selected stocks from Russell 3000.  The plotted is the
   the annualized volatility of the optimal portfolios by taking (a) the no short-sale
   portfolio as $Y$ and (b) the S\&P 500 stock index as $Y$.   The results are
   very similar and demonstrate that the optimal no-short-sale portfolio is
   not diversified whereas the global minimum portfolio is unstable.
   Both portfolios can be improved by an optimal portifolio with number of stocks
   around 100.}
\end{singlespace}
\end{center}
\end{figure}

We now apply our techniques to study the portfolio behavior using
Russell 3000 stocks.  The study period is from January 2, 2003 to
December 31, 2007. To avoid computation burden and the issues of
missing data, we picked 1000 stocks from 3000 stocks that
constitutes Russell 3000 on December 31, 2007. Those 1000 stocks
have least percents of missing data in the five-year study period.
This forms the universe of the stocks under our study. To mitigate
the possible survival biases, at the end of each month, we randomly
selected 400 stocks from the universe of the stocks. Therefore, the
400 stocks used in one month differs substantially from those used
in another month. The optimal no-short-sale portfolios, say, in one
month differ also substantially from that in the next month, because
they are constructed from very different pools of stocks.

At the end of each month from 2003 to 2007, the covariance of the
400 stocks is estimated according to various estimators using the
past 24 months' daily returns.  Since individual stocks have higher
volatility than individual portfolios, the longer time horizon than
that in the study of the 100 Fama-French portfolios is used. We use
these covariance matrices to construct optimal portfolios under
various gross-exposure constraints and hold these portfolios for one
month. The daily returns of these portfolios are recorded and hence
the standard deviations are computed.  We did not compute the mean
returns, as the universes of stocks to be selected from differ
substantially from one month to another, making the returns of the
portfolios change substantially from one month to another. Hence,
the aggregated returns are less meaningful than the risk.

Table 6 summarizes the risks of the optimal portfolios constructed
using 3 different methods of estimating covariance matrix and using
6 different gross-exposure constraints.  As the number of stocks
involved is 400, the quadratic programming package that we used can
fail to find the exact solution to problem (\ref{b6}).  It has too
many variables for the package to work properly.   Instead, we
computed only approximate solutions taking two different portfolios
as the $Y$ variable.

The global minimum portfolio is not efficient for vast portfolios
due to accumulation of errors in the estimated covariance matrix.
This can be seen easily from Figure 8.  The ex-post annualized
volatilities of constructed portfolios using the sample covariance
and RiskMetrics shoot up quickly (after 200 stocks chosen) as we
increase the number of stocks (or relax the gross-exposure
constraint) in our portfolio.  The risk continues to grow if we
relax further the gross-exposure constraint, which is beyond the
range of our pictures.  The maximum and minimum weights are very
extreme for the global minimum portfolio when the sample covariance
matrix and the RiskMetrics are used.   This is mainly due to the
errors in these estimated covariance matrices.  The problem is
mitigated when the gross-exposure constraints are imposed.

The optimal no-short-sale portfolios are not efficient in terms of
ex-post risk calculation.  They can be improved, when portfolios are
allowed to have 50\% short positions, say, corresponding to $c=2$.
This is due to the fact that the no-short-sale portfolios are not
diversified enough. The risk approximations are accurate beyond the
range of $c=1$.  On the other hand, the optimal no-short-sale
portfolios outperform substantially the global minimum portfolios,
which is consistent with the conclusion drawn in Jagannathan and Ma
(2003) and with our risk approximation theory.  When the
gross-exposure constraint is loose, the risk approximation is not
accurate and hence the empirical risk is overly optimistic.  As a
result, the allocation vector that we want from the true covariance
matrix is very different from the allocation vector that we get from
the empirical data.  As a result, the actual risk can be quite far
away from the true optimal.

The risks of optimal portfolios tend to be smaller and stable, when
the covariance matrix is estimated from the factor model.  For vast
portfolios, such an estimation of covariance matrix tends to be most
stable among the three methods that we considered here.  As a
result, its associated portfolio risks tend to be the smallest among
the three methods.   As the covariance matrix estimated by using
RiskMetrics uses a shorter time window than that based on the sample
covariance matrix, the resulting estimates tend to be even more
unstable.  As a result, its associated optimal portfolios tend to
have the highest risks.

The results that we obtain by using two different approximate
methods are actually very comparable.  This again provides an
evidence that the approximate algorithm yields the solutions that
are close to the exact solution.

\section{Conclusion}

The portfolio optimization with the gross-exposure constraint
bridges the gap between the optimal no-short-sale portfolio studied
by Jagannathan and Ma (2003) and no constraint on short-sale in the
Markowitz's framework. The gross-exposure constraint helps control
the discrepancies between the empirical risk which is always overly
optimistic, oracle risk which is not obtainable, and the actual risk
of the selected portfolio which is unknown.  We demonstrate that for
a range of gross exposure parameters, these three risks are actually
very close.  The approximation errors are controlled by the worst
elementwise estimation error of the vast covariance matrix. There is
no accumulation of estimation errors, thanks to the constraint on
the gross exposure.

We provided theoretical insights into the observation made by
Jagannathan and Ma (2003) that the optimal no-short-sale portfolio
has smaller actual risk than the global minimum portfolio for vast
portfolios and offered empirical evidence to strengthen the
conclusion.  We demonstrated that the optimal no-short-sale
portfolio is not diversified enough.  It is still a conservative
portfolio that can be improved by allowing some short positions.
This is demonstrated by our empirical studies and supported by our
risk approximation theory:  Increasing gross exposure somewhat does
not excessively increase the risk approximation errors, but
increases significantly the space of allowable portfolios and hence
decreases drastically the oracle risk and the actual risk.

Practical portfolio choices always involve constraints on individual
assets such as the allocations are no larger than certain
percentages of the median daily trading volume of an asset.  This is
commonly understood as an effort of reducing the risks of the
selected portfolios. Our theoretical result provides further
mathematical insights to support such a statement. The constraints
on individual assets also put a constraint on the gross exposure and
hence control the risk approximation errors, which makes the
empirical risk and actual risk closer.

Our studies have also important implications in the practice of
portfolio allocation.  We provide a fast approximate algorithm to
find the solution paths to the constrained risk minimization
problem.  We demonstrate that the sparsity of the portfolio
selection with gross-exposure constraint.  For a given covariance
matrix, we were able to find the optimal number of assets, ranging
from $N_0$ to the total number of stocks under consideration, where
$N_0$ is number of assets in the optimal no-short-sale portfolio.
This reduces an NP-complete hard optimization problem to a problem
that can be solved efficiently. In addition, the empirical risks of
the selected portfolios help us to select a portfolio with a small
actual risk. Our methods can also be used for portfolio tracking and
improvement.

\newpage

\begin{center}
\Large \bf Appendix A:  Conditions and Proofs
\end{center}

\renewcommand{\theequation}{A.\arabic{equation}}
\setcounter{equation}{0}

\bigskip

Throughout this appendix, we will assume that $\bmu = E \bR_t$ and
$\bS = E (\bR_t \bR_t^T)$ are independent of $t$. Let ${\cal F}_t$
be the filtration generated by the process $\{\bR_t\}$.
\bigskip

\noindent \textbf{Condition 1:} \footnote{The conditions are imposed
to facilitate the technical proof.  They are not weakest possible.
In particular, the condition such as  $\max_t E \| \beps_t \|_\infty
< \infty$ can be relaxed by replacing an upper bound depending on
$p$ such as $\log p$, and the conclusion continues to hold with some
simple modifications. The assumptions on matrices $\{\bA_j\}$ can
easily be checked when they are diagonal.  In particular, the
assumption holds when $\{\bR_t\}$ are a sequence of independently
identically distributed random vectors.}  Let $\bY_t$ be the
$p(p+1)/2$-dimensional vector constructed from the symmetric matrix
$\bR_t \bR_t^T - \bS$.  Assume that $\bY_t$ follows the vector
autoregressive model:
$$
  \bY_t = \bA_1 \bY_{t-1} + \cdots + \bA_k \bY_{t-k} + \beps_t,
$$
for coefficient matrices $\bA_1, \cdots, \bA_k$ with
$E\{\beps_t|{\cal F}_t\}=0$ and $\max_t E \| \beps_t\|_\infty <
\infty$.  Assume in addition that $\sup_{t} E \| \bA_j \bY_t
\|_\infty = O_p(n^{1/2})$ for all $1 \leq j \leq k$ and $\|
\bb_{(j)} \|_1 < \infty$ where $\bb_{(j)}$ is the $j$-th row of
matrix $\bB^{-1}$, with $\bB = \bI - \bA_1 - \cdots - \bA_k$.  In
addition, we assume similar conditions hold for the return vector
$\{\bR_t\}$.

\bigskip

Before introducing Condition 2, let us introduce the strong mixing
coefficient $\alpha(k)$ of the process $\{\bR_t\}$, which is given
by
$$
\alpha(k) = \sup_t   \sup\{|P(AB) - P(A) P(B)|: A \in \sigma(\bR_s, s\leq t), \;
B \in \sigma(\bR_s, s \geq t +k)\},
$$
where $\sigma(\bR_s, s\leq t)$ is the sigma-algebra generated by
$\{\bR_s, s\leq t\}$.

\bigskip

\noindent{\bf Condition 2}.   Suppose that $\| \bR_t\|_\infty < B$
for a constant $B > 0$ and  that as $q \to \infty$, $\alpha(q) =
O(\exp(- C q^{1/b}))$ and $a > (b+1)/2$ in Theorem 3.  In addition,
$\log n = O(\log p)$. \footnote{In
the case that $n$ is very large so that $\log n$ is of a larger
order than $\log p$, the conclusion still holds with $\log p$ in
Theorem 3 replaced by $\log n$.}

\bigskip

\noindent{\bf Condition 3}.  Let $\eta_t$ be $R_{ti} R_{tj} -
ER_{ti}R_{tj}$ or $R_{ti} - ER_{ti}$ (We suppress its dependence on
$i$ and $j$).  Assume that  for all $i$ and $j$ there exist
nonnegative constants $a$, $b$, and $B$ and a function $\rho(\cdot)$
such that
$$
    |\cov(\eta_{s_1} \cdots \eta_{s_{u}}, \eta_{t_1} \cdots \eta_{t_v})| \leq
    B^{u+v} [(u+v)!]^b v \rho(t_1 - s_u),
$$
for any $1 \leq s_1 \leq \cdots \leq s_u \leq t_1 \leq \cdots \leq
t_v \leq n$ where
$$
    \sum_{s=0}^\infty (s+1)^k \rho (s) \leq B^k (k!)^a
    \qquad \mbox{for all $k > 0$.}
$$
and \footnote{Neumann and Paraporodidis (2008) show that this
covariance weak dependence condition holds for AR and ARCH processes
with $a = 1$, $b=0$ and $\rho(s) = h^s$ for some $h < 1$.}
$$
    E |\eta_t|^k \leq (k!)^\nu B^k, \qquad \mbox{for all $k > 0$.}
$$
In addition, we assume that $\log p = o(n^{1/(2a + 2 b +3)})$.

\bigskip

\noindent {\em Proof of Theorem 1}:  First of all, $R(\hat
\bw_{opt}) - R(\bw_{opt}) \geq 0$, since $\bw_{opt}$ minimizes the
function $R$.  Similarly, we have $R_n(\hat
\bw_{opt})-R_n(\bw_{opt})\leq 0$.  Consequently, we have
\begin{eqnarray}
        R(\hat \bw_{opt}) - R(\bw_{opt})
  & = & R(\hat \bw_{opt}) - R_n(\hat \bw_{opt}) + R_n(\hat \bw_{opt})
         -R_n(\bw_{opt})+R_n(\bw_{opt})-R(\bw_{opt})\nonumber\\
  & \leq & R(\hat\bw_{opt})-R_n(\hat \bw_{opt})+R_n(\bw_{opt})-R(\bw_{opt})\nonumber\\
  & \leq & 2 \mbox{sup}_{||\bw||\leq c}|R_n(\bw)-R(\bw)|. \label{A1}
\end{eqnarray}
Now, it is easy to see that
\begin{equation}
 |R_n(\bw)-R(\bw)| = | \bw^T ( \hat{\bSigma} -\bSigma) \bw | \leq a_n \|\bw\|_1^2, \label{A2}
\end{equation}
which is bounded by $a_nc^2$.  This together with (\ref{A1}) proves
the first conclusion and the second conclusion.

To prove the third inequality, we note that
\begin{eqnarray*}
        |R(\bw_{opt})-R_n(\hat{\bw}_{opt})|
    &\leq & |R(\bw_{opt})-R(\hat{\bw}_{opt}) |
            + |R(\hat\bw_{opt})-R_n(\hat{\bw}_{opt}) | \nonumber\\
& \leq & 3 \mbox{sup}_{||\bw||\leq c} |R_n(\bw)-R(\bw)|
\end{eqnarray*}
where we used (\ref{A1}) to bound the first term. The third
inequality comes from (\ref{A2}).

\bigskip

We need the following lemma to prove Theorem 2.

\bigskip

\textbf{Lemma 1}. Let $\bxi_1, \cdots, \bxi_n$ be a $p$-dimensional
random vector. Assume that $\bxi_t$ is $\mathcal{F}_t$-adaptive and
each component is a martingale difference:
$E(\bxi_{t+1}|\mathcal{F}_t)=0$. Then, for any $p\geq 3$ and
$r\in[2,\infty]$, we have for some universal constant $C$
 \begin{equation} \label{A3}
 E||\sum_{t=1}^{n}\bxi_t||_r^2\leq C \: \mbox{min}[r, \log p]
 \sum_{t=1}^{n}E||\bxi_t||_r^2
 \end{equation} where $||\bxi_t||_r$ is the $l_r$-norm of the
vector $\bxi_t$ in $R^p$.

\bigskip

This is an extension of the Nemirovski's inequality to the marginale
difference sequence.  The proof follows similar arguments on page
188 of Emery el al (2000).

\noindent {\em Proof of lemma 1}.  Let $V(\bx)=||\bx||_r^2$. Then,
there exists a universal constant $C$ such that
$$
  V(\bx+\by)\leq V(\bx)+\by^T V'(\bx) + C r V(\by),
$$
where $V'(\bx)$ is the gradient vector of $V(\bx)$. Using this, we
have
\begin{equation}
  V(\sum_{t=1}^{n}\bxi_t)\leq
  V(\sum_{t=1}^{n-1}\bxi_t)+\bxi_n^T V'(\sum_{t=1}^{n-1}\bxi_t) + C rV(\bxi_n).
  \label{A4}
\end{equation}
Since $\bxi_n$ is a martingale difference and
$V'(\sum_{t=1}^{n-1}\bxi_t)$ is $\mathcal{F}_{n-1}$ adaptive, we
have
$$
    E \bxi_n^T V'(\sum_{t=1}^{n-1}\bxi_t)=0.
$$
By taking the expectation on both sides of (\ref{A4}), we have
$$
    E V(\sum_{t=1}^{n}\bxi_t) \leq E V (\sum_{t=1}^{n-1}\bxi_t) + CrE V(\bxi_n).
$$
Iteratively applying the above formula, we have
\begin{equation}
    E \|\sum_{t=1}^{n}\bxi_t\|_r^2 \leq C r\sum_{t=1}^{n} \|\bxi_t\|_r^2.  \label{A5}
\end{equation}
This proves the first half of the inequality (\ref{A3}).

To prove the inequality (\ref{A3}), without loss of generality,
assume that $r\geq \log p.$ Let $r'=\log p > 1$. Then, for any $\bx$
in the $p$-dimensional space,
$$
  \|\bx\|_r \leq \|\bx\|_{r'}\leq
p^{\frac{1}{r'}-\frac{1}{r}}||\bx||_r
$$
Hence, by (\ref{A5})
\begin{eqnarray*}
E||\sum_{t=1}^{n}\xi_t||_r^2 & \leq & E||\sum_{t=1}^{n}\bxi_t||_{r'}^2\nonumber \\
& \leq & C \log p \sum_{t=1}^{n}E||\bxi_t||_{r'}^2\nonumber\\
& \leq & C \log p
\sum_{t=1}^{n}p^{2(\frac{1}{r'}-\frac{1}{r})}E||\bxi_t||_r^2
\end{eqnarray*}
Using the simple fact $p^{\frac{2}{r'}}=e^2$, we complete the proof
of the inequality (\ref{A3}).

\bigskip

\noindent{\em Proof of Theorem 2}.  Applying lemma 1, with
$r=\infty$, we have
\begin{equation}
    E  \| n^{-1} \sum_{t=1}^{n}\bxi_t\|_\infty^2
    \leq \frac{C \log p}{n}  \max_t E \|\bxi_t\|_{\infty}^2, \label{A6}
\end{equation}
for all $t$, where $E \|\bxi_t\|_{\infty}^2 = E (\max_{1 \leq j \leq
p} \xi_{tj}^2)$.  As a result, by Condition 1, an application of
(\ref{A6}) to $p(p+1)/2$-element of $\beps_t$ yields
$$
E  \| (n-k)^{-1} \sum_{t=k+1}^{n} (\bY_t - \bA_1 \bY_{t-1} - \cdots
- \bA_k \bY_{t-k} )\|_\infty^2
    \leq \frac{C \log p^2}{n-k} \max_t E \|\beps_t \|_{\infty}^2.
$$
Note that each of the summation $(n-k)^{-1} \sum_{t=k+1}^n
\bY_{t-j}$  (for $j \leq k$) is approximately the same as $n^{-1}
\sum_{t=1}^{n}\bY_t$ since $k$ is finite, by appealing to Condition
1.  Hence, we can easily show that
$$
\| \bB n^{-1} \sum_{t=1}^{n}\bY_t \|_\infty = O_p\left
(\sqrt{\frac{\log p}{n}} \right ).
$$
By the assumption on the matrix $\bB$, we can easily deduce that
$$
\| n^{-1} \sum_{t=1}^{n}\bY_t \|_\infty  = O_p\left
(\sqrt{\frac{\log p}{n}} \right ).
$$
Rearranging this into matrix form, we conclude that
$$
   \| n^{-1} \sum_{t=1}^{n}\bR_t \bR_t^T - \bS \|_\infty =  O_p\left
(\sqrt{\frac{\log p}{n}} \right ).
$$

Let $\bar{\bR}_n =  n^{-1}  \sum_{t=1}^{n}\bR_t$.  As $\bR_t$
satisfies similar conditions to $\bY_t$, we have also that
$$
  \| \bar{\bR}_n - \bmu \|_\infty = O_p \left (
    \sqrt{\frac{\log p}{n}} \right ).
$$

Finally, by using
$$
     \hat{\bS}_n = n^{-1} \sum_{t=1}^{n}\bR_t \bR_t^T - \bar{\bR}_n \bar{\bR}_n^T,
$$
we conclude that
$$
  \| \hat{\bS}_n - \bSigma \|_\infty = O_p\left (\sqrt{\frac{\log p}{n}} \right ).
$$

\bigskip

\noindent {\em Proof of Theorem 3}. Note that by the union bound of
probability,  we have for any $D > 0$,
$$
P\{\sqrt{n} \|\bSigma - \hat{\bSigma} \|_\infty > D (\log p)^{a}\}
\leq p^2 \max_{i, j} P\{ \sqrt{n} | \sigma_{ij} - \hat{\sigma}_{ij}|
> D (\log p)^{a} \}.
$$
By the assumption, the above probability is bounded by
$$
   p^2 \exp\left (- C [D (\log p)^a]^{1/a} \right ) = p^2 p^{-CD^{1/a}},
$$
which tends to zero when $D$ is large enough.  This proves the first
part of the theorem.

We now prove the second part of the $\alpha$-mixing process.  Let
$\xi_t$ be an $\mathcal{F}_t$ adaptive random variable with $E \xi_t
=0$ and assume that $|\xi_t| \leq B$ for all $t$.  Then, by Theorem
1.3 of Bosq (1998), for any integer $q \leq n/2$, we have
\begin{equation}
   P( |\bar{\xi}_n| > \varepsilon )
\leq 4 \exp(- \frac{q \varepsilon^2}{8B^2}) + 22 (1 +
4B/\varepsilon)^{1/2}
       q \alpha \left ( [n/(2q)] \right ),  \label{A7}
\end{equation}
where $\bar{\xi}_n = n^{-1} \sum_{t=1}^n \xi_t$.  Taking
$\varepsilon_n =4 B D (\log p)^a / \sqrt{n}$ and $q = n (\log
p)^{1-2a}/2$, we obtain from (\ref{A7}) that
$$
P( |\bar{\xi}_n| > \varepsilon_n ) = 4 p^{-D^2} + o(n^{3/2}) \alpha( (\log p)^{2a -1})
$$
Now, the assumption on the mixing coefficient $\alpha(\cdot)$, we
conclude that for sufficiently large $D$,
\begin{equation}
  P( |\bar{\xi}_n| > \varepsilon_n ) = o(p^{-2}),   \label{A8}
\end{equation}
for $a > (b+1)/2$.

Applying (\ref{A8}) to $\xi_t = R_{ti} R_{tj} - E R_{ti} R_{tj}$
with a sufficiently large $D$, we have
$$
 P( n^{-1} \sum_{t=1}^n | R_{ti} R_{tj} - E R_{i}
R_{j}| > \varepsilon_n ) = o(p^{-2}).
$$
This together with the first part of proof of Theorem 2 yield that
$$
   \| n^{-1} \sum_{t=1}^{n}\bR_t \bR_t^T - \bS \|_\infty =  O_p\left (
\varepsilon_n  \right ),
$$
where we borrow the notation from the proof of Theorem 2. Similarly,
by an application of (\ref{A8}), we obtain
$$
 \| \bar{\bR}_n - \bmu \|_\infty = O_p\left (
\varepsilon_n  \right ).
$$
Combining the last two results, we prove the second part of the
theorem.

The proof of the third part of the theorem follows similar steps.
By Theorem 1 of Doukhan and Neumann (2007), under Condition 3, we
have
$$
P( |\sum_{t=1}^n \eta_t | > \sqrt{n} x )  \leq \exp(- C\min\{x^2,
(\sqrt{n}x)^c\})
$$
for some $C > 0$, where $c = 1/(a+b+2)$.  Now, taking $x = D (\log
p)^{1/2}$, we have
$$
   x^2/(\sqrt{n}x)^c = O( (\log p)^{1-c/2} / n^{c/2}) = o(1),
$$
since $\log p = o(n^{1/(2\mu + 2 \nu +3)})$.  Thus, the exponent is
as large as
$$
 C \min\{x^2,(\sqrt{n}x)^c\} \geq CD^2 \log p,
$$
for sufficiently large $n$. Consequently,
$$
    P( |\sum_{t=1}^n \eta_t | > D \sqrt{n \log p} ) \leq \exp(-CD^2
    \log p) = o(p^{-2})
$$
for sufficiently large $D$.  Now, substituting the definition of
$\eta_t$, we have
\begin{eqnarray}
& &P( n^{-1} \sum_{t=1}^n | R_{ti} R_{tj} - E R_{i} R_{j}| > D
\sqrt{(\log p)/n} ) = o(p^{-2}).   \label{A9} \\
& & P( n^{-1} \sum_{t=1}^n | R_{ti} - E R_{i}| > D \sqrt{(\log p)/n}
) = o(p^{-2}).   \label{A10}
\end{eqnarray}
Combining the results in (\ref{A9}) and (\ref{A10}) and using the
same argument as proving the first part of Theorem 2, we have
$$
 \| n^{-1} \sum_{t=1}^{n}\bR_t \bR_t^T - \bS \|_\infty =  O_p\left (
   \sqrt{(\log p)/n}  \right ).
$$
and
$$
  \| n^{-1} \sum_{t=1}^{n}\bR_t  - \bmu \|_\infty =  O_p\left (
   \sqrt{(\log p)/n}  \right ).
$$
The conclusion follows from these two results.


\newpage

\bigskip

\begin{center}
\Large \bf Appendix B:  LARS-LASSO Algorithms for Constrained Risk Minimization
\end{center}

\renewcommand{\theequation}{B.\arabic{equation}}
\setcounter{equation}{0}

\bigskip

We now describe the LARS-LASSO algorithm for the constrained
least-square problem ({\ref{c3}). First, standardize each variable
$X_j$ so that it has unit variance.  The basic idea is very
intuitive. As soon as $d$ moves slightly away from zero, one picks
only one variable, which has the maximum absolute correlation with
the response variable $Y$. Without loss of generality, let us assume
that the maximum absolute correlation achieves at the first variable
and the correlation is negative. Then, $\bw^* = (-d, 0, \cdots,
0)^T$ is the solution to problem (\ref{c3}) for some small $d$. Now,
as $d$ increases, the absolution correlation of the working residual
$R = Y - \bX^T \bw  = Y + d X_1$ with $X_1$ decreases until a
(smallest) value $d_1$ at which there exists a second variable
$X_2$, say, that has the same absolution correlation with $R$. Then,
$\bw$ is the solution to problem (\ref{c3}) for $0 \leq d \leq d_1$.
For $d$ slightly bigger than $d_1$, there are two non-vanishing
components in $\bw_\delta = \bw_1 + \delta \bgamma$, where $\bw_1 =
(-d_1, 0, \cdots, 0)$ and the direction $\bgamma$, having only first
two elements non-vanishing, is chosen so that the absolute
correlations of the working residual $R = Y- \bX^T \bw_\delta$ with
$X_1$ and $X_2$ decrease equally as $\delta$ increases until a point
$\delta_1$ at which a third variable, $X_3$, say, has the same
absolute correlation with the working residual as those  with $X_1$
and $X_2$.  The solution to problem (\ref{c3}) simply $\bw_\delta$
for $d \in (d_1, \|\bw_{\delta_1}\|_1]$. For $d$ going slightly
beyond that point, the solution to problem (\ref{c3}) consists of 3
variables. Continuing this process, we will get the whole solution
path.

The LARS algorithm runs as follows.  Let
$$
    \sigma_j = \cov(X_j, Y) \quad \mbox{and} \quad
    \bxi_j   = \cov(X_j, \bX),
$$
which are obtained from the input covariance matrix.
\begin{enumerate}
\item Set the initial value $\bw = 0$.  This corresponds to the solution with $d=0$.

\item Compute $u = \max_j | \sigma_j - \bxi_j ^T \bw|$, which is the
maximum absolute correlation (multipled by the standard deviation of
$Y$) between the working residual $R = Y - \bX^T \bw$ and $X_j$. Let
$\calC$ be index of assets that achieved the maximum absolute
correlation.

\item Increase the value $\bw$ for the components in $\calC$ in the direction
$\bgamma_{\calC}$ until a new variable is added to the set $\calC$.
The direction $\bgamma_{\calC}$ is chosen so that the absolute
correlations of the working residual with all variables $\{ X_j,
j\in \calC\}$ decrease equally. The direction $\bgamma_{\calC}$ can
easily be determined analytically and so is the thresholding value
for the amount of the increase.

\item Repeat Steps 2 and 3 until all variables are recruited.
\end{enumerate}

\end{document}